\documentclass[fleqn,usenatbib]{mnras}

\usepackage{newtxtext,newtxmath}

\usepackage[utf8]{inputenc}
\usepackage[T1]{fontenc}

\DeclareRobustCommand{\VAN}[3]{#2}
\let\VANthebibliography\thebibliography
\def\thebibliography{\DeclareRobustCommand{\VAN}[3]{##3}\VANthebibliography}

\usepackage{graphicx}	
\usepackage{amsmath}	
\usepackage[normalem]{ulem} 

 \def\Mauro#1{{\textcolor{black}{#1}}}

\newcommand{\SFmodel}{blue}
\newcommand{\GTmodel}{black}
\newcommand{\Potmin}{red}
\newcommand{\Ntot}{1997} 

\newcommand{\GT}{GT}
\newcommand{\LMV}{LMV$_z$}
\newcommand{\SF}{SF}

\newcommand{\kms}{km s$^{-1}$}
\newcommand{\kmskpc}{km s$^{-1}$ kpc$^{-1}$}

\title[Time Evolution of the Warp]{Structure, kinematics and time evolution of the Galactic Warp from Classical Cepheids}

\author[M. Cabrera-Gadea et al.]{
Mauro Cabrera-Gadea$^{1}$\thanks{E-mail: mauro.cabrera@pedeciba.edu.uy},
Cecilia Mateu$^{1}$,
Pau Ramos$^{2}$,
Mercè Romero-G\'omez$^{3,4,5}$,
Teresa Antoja$^{3,4,5}$,\newauthor
and Luis Aguilar$^{6}$
\\
$^{1}$Departamento de Astronom\'{i}a, Instituto de F\'{i}sica, Universidad de la Rep\'{u}blica, Igu\'a 4225, CP 11400 Montevideo, Uruguay \\
$^{2}$National Astronomical Observatory of Japan, Mitaka-shi, Tokyo 181-8588, Japan\\
$^{3}$Institut de Ci\`encies del Cosmos (ICCUB), Universitat de Barcelona (UB), Mart\'{i} i Franqu\`es 1, E-08028 Barcelona, Spain\\
$^{4}$Departament de F\'{i}sica Qu\`antica i Astrof\'{i}sica (FQA), Universitat de Barcelona (UB), Mart\'{i} i Franqu\`es 1, E-08028 Barcelona, Spain\\
$^{5}$Institut d'Estudis Espacials de Catalunya (IEEC), c. Gran Capit\`a, 2-4, E-08034 Barcelona, Spain.\\
$^{6}$ Instituto de Astronom\'{i}a, Universidad Nacional Auton\'{o}ma de M\'{e}xico, Apdo. Postal 877, Ensenada, 22800 Baja California, Mexico
}
\date{Accepted XXX. Received YYY; in original form ZZZ}

\pubyear{2023}

\begin{document}
\label{firstpage}
\pagerange{\pageref{firstpage}--\pageref{lastpage}}
\maketitle

\begin{abstract}
The warp is a well-known undulation of the Milky Way disc. Its structure has been widely studied, but only since Gaia DR2 has it been possible to reveal its kinematic signature beyond the solar neighbourhood. In this work we present an analysis of the warp traced by Classical Cepheids by means of a Fourier decomposition of their height ($Z$) and, for the first time, of their vertical velocity ($V_z$). We find a clear but complex signal that in both variables reveals an asymmetrical warp. In $Z$ we find the warp to be almost symmetric in amplitude at the disc's outskirts, with the two extremes never being diametrically opposed at any radius and the line of nodes presenting a twist in the direction of stellar rotation for $R>11$ kpc. For $V_z$, in addition to the usual $m=1$ mode, an $m=2$ mode is needed to represent the kinematic signal of the warp, reflecting its azimuthal asymmetry. The line of maximum vertical velocity is similarly twisted as the line of nodes and trails behind by $\approx 25^\circ$. We develop a new formalism to derive the pattern speed and change in amplitude with time $\dot{A}$ of each Fourier mode at each radius, via a joint analysis of the Fourier decomposition in $Z$ and $V_z$. By applying it to the Cepheids we find, for the $m=1$ mode, a constant pattern speed in the direction of stellar rotation of $9.2\pm3.1$ km/s/kpc, a negligible $\dot{A}$ up to $R\approx 14$ kpc and a slight increase at larger radii, in agreement with previous works.
\end{abstract}




\section{Introduction}

The warp is an undulation in a galactic disc that makes its mean vertical height deviate from the mid plane in the outskirts of the galaxy. 
Between $40-50\%$ of edge-on disc galaxies are found to be warped \citep{Sanchez-Saavedra_1990_Statistical_Study_of_Optical_Galactic_WARPS,Reshetnikov_Statistics_of_optical_WARPS_1998}, which implies that warps should be long lived phenomena or the formation mechanism a very recurrent one in the history of galactic discs. The Milky Way is not an exception, having a warp whose structure has been widely studied with different tracers like HI \citep{Levine2006}, dust \citep{Marshall_Modelling_interstellar_extinction_2006} as well as with different stellar populations \citep{Warp_RedClump_Lopez_Corredoira_2002,RG19,Skowron2020,Chen3Dmap,Cheng_2020_Warp_patern,Chrobacova_GDR2_Warp_Model_2020GDR2,Li_2023_Nonsteady_Warp}. Although the Galactic warp has been known for a long time \citep{Burke_HI_Warp_1957}, its origin is still puzzling. In order to elucidate the history and formation of the Milky Way's warp, it is important to characterise its main properties, as its structure and kinematics.

Classical Cepheids have proven exceptionally useful in tracing the structure and kinematics of the warp offering several key advantages to study the Galactic disc \citep{Bobylev2013_warp,Skowron2019,Skowron2020,Chen3Dmap}. 
Being very young stars \citep[with ages up to a few hundred million years, e.g.][]{CatelanSmith2015} it is expected that they have recently inherited the warped structure of the HI gas where they have formed, while still having cold kinematics (vertical velocity dispersion $<5$\kms, \citealt{Chen3Dmap}) making it easier to observe the warp signal as secular dynamics has not had time to 'heat' or disturb it, as it would have for older populations \citep[Sec. 8.1]{BT08}. Also, belonging to such a young population means there is no contamination from any other Galactic component, e.g. the thick disc or halo, which means they exclusively trace the Galactic thin disc. In addition,  Classical Cepheids are well-known standard candles \citep{Leavitt1908,Leavitt1912}, offering extremely precise distance measurements ($\sim3\%$ errors); they can be reliably identified based on their variability, making contamination from other stars negligible \citep[e.g.][]{Jayasinghe2019,Rimoldini2019,Rimoldini2023}; and being luminous ($500<L/L_\odot<20,000$, \citealt{CatelanSmith2015}), makes them observable throughout a large extent of the disc even with the optical surveys used to identify them at present \citep{Udalski_OGLE_Cepheids_2018,Ripepi2023_GaiaDR3Cepheids}. Their only disadvantage is that they are relatively scarce, with fewer than $2500$ Classical Cepheids in the deepest and most complete catalogues of the Galactic disc to date provided by OGLE-IV \citep{Udalski_OGLE_Cepheids_2018} and the Third Gaia Data Release \citep[DR3,][]{GaiaDR3,Ripepi2023_GaiaDR3Cepheids}. For these reasons, Cepheids have been used to study the 3D structure of the warp in more detail than any other stellar population \citep{Skowron2019}, providing evidence for a twisted line of nodes \citep{Chen3Dmap,Dehnen_2023_Warp_Ceph} and asymmetry in height between both extremes \citep{Skowron2020} similar to the HI warp \citep{Levine2006}. They have also been used to study the kinematics of the warp revealing its characteristic bulk vertical motion in the outskirts of the disc \citep{Skowron2020}.

To describe the structure of the warp, several studies have shown Fourier series to be of great use due to their versatility to summarise any warp signal if enough modes are considered \citep[e.g.][]{Levine2006,Chen3Dmap,Skowron2020}. These studies have focused on describing the structure of the warp, i.e. its mean height as a function of radius and azimuth. Using a catalogue of Classical Cepheids identified mainly with OGLE-IV and combined with Gaia DR2 astrometry, \citet{Skowron2019} used a Fourier decomposition with up to 2 modes ($m\leq2$) and a fixed line of nodes (LON) to present the first map of the Galactic warp in the young population covering over half the disc. \citet{Chen3Dmap}, using a compilation of optical plus Wide-field Infrared Survey Explorer \citep[WISE,][]{Chen_WISE_2018} Cepheid catalogues, studied the azimuthal dependence of the LON with radius finding it does not coincide with the Sun-Galactic Centre direction and that it presents a leading pattern, following Briggs's rules for HI warps in spiral galaxies \citep{Briggs_Rules_for_Galactic_WARPS}. For the kinematics, Fourier series have been used to characterise the changes in mean vertical velocity ($V_z$) in simulations \citep{Chequers2018,Poggio_Measuring_vertical_response_2021}, but insofar there have been no Fourier decomposition studies of the warp's kinematic signal with Cepheids (or any tracer) which can reflect and quantify its plausible azimuthal asymmetries and changes with radius. Previous studies with other --older-- stellar populations have assumed the kinematic signal to be well represented by an $m=1$ mode \citep{Poggio_Warp_evolving,Cheng_2020_Warp_patern,Wang_LAMOST_Disk_Warp_RC_2020,Chrobakova_Against_wapr_Precession_2021,Dehnen_2023_Warp_Ceph}, as expected from a tilted rings model ($m=1$ mode), but \cite{RG19} argue this model is insufficient to explain the more complex kinematic signature they observed with Red Clump stars. 

In this work we use a Fourier Decomposition method to study the structure \emph{and kinematics} of the Galactic warp using Classical Cepheids as tracers. We use the Cepheid catalogue from \citet{Skowron2019} combined with kinematic data from Gaia DR3 \citep{GaiaDR3} to explore the dependence of the amplitudes and the azimuths of the modes as free parameters as a function of radius, which  allows us to infer the position of the LON and Line of Maximum Vertical Velocity (\LMV) for a general warp model that accounts for the lopsidedness of the warp. The new method we present here (Sec.~\ref{sec:Disc_TimeEvol}), based on a joint analysis of the Fourier series for $Z$ and $V_z$, allows us to infer the time evolution of the Fourier components of the warp: i.e. their pattern speed and instantaneous change in amplitude. The inference of the evolutionary terms of the Galactic warp has been tackled recently using different tracers, but mostly under the tilted rings model which assumes a symmetric warp. \citet{Poggio_Warp_evolving} and \citet{Cheng_2020_Warp_patern}  have focused on inferring the pattern speed, while \citet{Wang_LAMOST_Disk_Warp_RC_2020} derived the change in amplitude. These works use general samples of stars with good quality Gaia DR2 and DR3 astrometry and available radial velocities; by not been focused on a specific tracer or having any age constraints, these parameters are representative of the general population of the disc as weighed by its star formation history, i.e. a stellar population of intermediate age (several Gyrs old). The recent work by \citet{Dehnen_2023_Warp_Ceph} derives both evolutionary terms for different radii for a sample of Cepheids via a tilted rings model, finding differential rotation and change in inclination of the rings. Our work uses the same stellar population to derive the time evolution parameters of the warp with a completely independent method. 

The structure of the present paper is as follows. In Section~\ref{sec:FDM} we present the Fourier decomposition method used to describe the Galactic warp's height and vertical velocity (Sec.~\ref{sec:fourier_zvz}) and the method showing how these are combined to derive each mode's pattern speed and amplitude change  (Section~\ref{Sec:AdotOmega}). In Section~\ref{s:inference} we present the inference model used to estimate the warp model's parameters, including main conclusions from the inference validation performed using a mock catalogue. In Sec.~\ref{sec:Sample} we describe the catalogue of Classical Cepheids used in this work. 
In Section~\ref{sec:ResultsFULL} we apply the methods to this sample and summarise our results for the structure and kinematics of the Cepheid's warp (Sec.~\ref{sec:Results}) and those for the time evolution (Section~\ref{Sec:TimeEvol}). 
In Section~\ref{sec:Discussion} we discuss our results and compare with the previous literature. Our conclusions are summarised in Section~\ref{sec:conclusions}. 

\section{Fourier Decomposition Method}\label{sec:FDM}

\subsection{Reference Frame}

We begin by describing the coordinate system and reference frame we use throughout this paper. The origin of the reference frame is at the galactic center (GC), fixed with respect to an external inertial frame. Positions can be given in Cartesian, or cylindrical coordinates. The X-axis points from the GC away from the Sun, the Y-axis is parallel to the rotation velocity of the disc at the Sun position and the Z-axis is perpendicular to the Galactic plane forming a right-handed triad. In cylindrical coordinates we use the Galactocentric azimuthal angle $\phi$ measured from the X-axis toward the Y-axis (i.e. opposite to Galactic rotation). In this coordinate system the Sun is at $R_\odot=8.277$ kpc \citep{GRAVITY_collab_2022_Rsun}, $\phi_\odot=180^\circ$ and $Z_\odot=0.027$ kpc \citep{ZsolarChen}. For velocities we use a Cartesian system whose origin is at rest with the GC and their axes parallel to the directions in which the X-Y-Z axes increase. This is an inertial system and thus does not rotate with the Galaxy, the Sun being along the negative X-axis only at present. This facilitates the kinematical and dynamical descriptions. We assume the Sun has Galactocentric cartesian velocity $(V_x,V_y,V_z)=(11.10, 232.24, 7.25)~\text{km/s}$ \citep{V_sun_Schonrich_2010,Bovy_Astropy_2015}. 

\subsection{Fourier Decomposition of the Structure and Kinematics}\label{sec:fourier_zvz}

We implement the Fourier decomposition method following  \cite{Levine2006} and \cite{Chequers2018}. The disc is divided into concentric Galactocentric rings, in each ring the mean behaviour as a function of the azimuth for $Z$ and $V_z$ described by a Fourier sums up to $M$ modes as

\begin{equation}
    Z(\phi)=\sum_{m=0}^M A_m\sin(m\phi-\varphi_m) 
    \label{Eq:2.Fourier_Aphi_Z} \\
\end{equation}

\begin{equation}
    V_z(\phi)=\sum_{m=0}^M V_m\sin(m\phi-\varphi^V_m).
    \label{Eq:2.Fourier_Aphi_Vz}
\end{equation}

The amplitudes $(A_m,V_m)$ and phases $(\varphi_m,\varphi_m^V)$ are free parameters, obtained as a function of $R$. In what follows we describe the method only for $Z(\phi)$, being analogous for $V_z(\phi)$. The Fourier representation is flexible enough, even with a small number of modes, to describe many known warp shapes: for example a U-shaped warp will be mainly described by an $m=0$ mode with increasing amplitude as a function of radius; an integral or S-shaped warp, will be mainly described by an $m=1$ mode with increasing amplitude as a function of radius, and asymmetries will be mainly described by a combination of $m=1$ and $m\geq 2$ modes. 

As we will see in Sec~\ref{s:inference}, it is also convenient to rewrite Eq.~\ref{Eq:2.Fourier_Aphi_Z} in linear form with free parameters $a_m,b_m$ as 

\begin{equation}
    Z(\phi)=\sum_{m=0}^M a_m\cos(m\phi)+b_m\sin(m\phi),
    \label{Eq:2.Fourier_AnBn}
\end{equation}

\noindent where the transformation between $a_m,b_m$ and $A_m,\phi_m$ is given by

\begin{equation}
    A_m  =  \sqrt{a_m^2+b_m^2} \ \ \ ; \ \ \   \varphi_m  =  \text{arctan2}(-a_m,b_m).
    \label{Eq:2.Transf}
\end{equation}

\subsection{Deriving Time Evolution}\label{Sec:AdotOmega}

In this section we present a new formalism to derive the evolutionary terms from the warp, its pattern speed and the change in amplitude of each mode at each radii, disentangled from the motion of the stars. From now on we denote the star's vertical height and vertical velocity as $z$ and $v_z$ (lowercase) and the Fourier fits to the warp as $Z$ and $V_z$ (uppercase).

We begin by taking a ring at a radius $R$ and considering a star that has no radial motion and constant angular velocity $\Omega$, that it simply rotates around the Galactic Centre but following the warp of a razor thin disc. These assumptions are reasonable for dynamically cold populations such as the Cepheid stars we use in our analysis. Given that the stars follow the warp's shape, their height $z(t)$ at time $t$  is given by the functional expression of the warp $Z(\phi,t)$, which we can express as a Fourier series evaluated at the star's azimuth $\phi(t)$ as follows

\begin{equation}
    z(t)=Z\Big(\phi(t),t\Big)=\sum_{m=0}^M A_{m}(t)\sin(m\phi(t)-\varphi_{m}(t)).
    \label{Eq:5.Z_t}
\end{equation}

 We allow the amplitude and phase of each mode to evolve in time because we are interested in determining their instantaneous derivatives $\dot{A}_m$ and $\dot{\varphi}_m$. If we take the total derivative of $z(t)$  with respect to time we obtain the vertical velocity $v_z$ of the \emph{star} --not the warp-- given by 

\begin{multline}
v_z(t) = \sum_{m=0}^M\dot{A}_{m}(t)\sin(m\phi(t)-\varphi_{m}(t))+ \\
     A_{m}(t)\cos(m\phi(t)-\varphi_{m}(t))[m\dot{\phi}(t)-\dot{\varphi}_{m}(t)]. \label{Eq:5.Vz_t}
\end{multline}

As expected, Eq.~\ref{Eq:5.Vz_t} involves terms regarding the time evolution of the warp ($\dot{A}_m$ and $\dot{\varphi}_m$), and a term regarding the motion of the star due to its own angular velocity ($\dot{\phi}(t)=\Omega$). Now we want to link Eq.~\ref{Eq:5.Vz_t}, which describes the velocity of just one star at azimuth $\phi(t)$, to the $Z(\phi)$ and $V_z(\phi)$ fits from the previous section, which describe the mean motion of all stars in the ring at a given time $t_0$ (today).

In a razor thin disc the height of the disc at an arbitrary azimuth and the position of a star at the same azimuth must exactly coincide. 
Thus, it follows that the vertical height $z(t_0)$ and vertical velocity $v_z(t_0)$ of a star at $t_0$ and azimuth $\phi(t_0)=\phi_0$ must coincide with the Fourier fits ($Z(\phi_0),V_z(\phi_0)$)  we obtained at that same $t_0$ time. Taking $t_0$ as today, $A_m(t_0)=A_{m,0}$ and $\varphi_m=\varphi_{m,0}$, the amplitudes and phases obtained from the Fourier fits from Eqs.~\ref{Eq:2.Fourier_Aphi_Z} and \ref{Eq:2.Fourier_Aphi_Vz}. 

Similarly, the vertical velocity $v_z(t_0)$ of the \emph{star} must also coincide with the mean vertical velocity obtained from our Fourier fit $V_z(\phi_0)$, evaluated at the star's azimuth. Setting $V_z(\phi_0)=v_z(t_0)$ in the left hand side of Eq.~\ref{Eq:5.Vz_t} and expressing $V_z(t_0)$ as the Fourier fit for $v_z$ in its linear form (as shown in Eq.~\ref{Eq:2.Fourier_AnBn} for $Z$) we obtain 

\begin{multline}
\sum_{m=0}^M a^V_m\cos(m\phi_0)+b^V_m\sin(m\phi_0)=\sum_{m=0}^M\dot{A}_{m}(t_0)\sin(m\phi_0-\varphi_{m})+ \\
A_{m}\cos(m\phi_0-\varphi_{m})[m\Omega-\dot{\varphi}_{m}(t_0)],
    \label{Eq:WarpEv.vzVZ}
\end{multline}

\noindent where $a^V_m$ and $b^V_m$ are the linear amplitudes resulting from the Fourier fits in velocity, calculated from the $V_m$ and $\varphi^V_m$ obtained via Eq.~\ref{Eq:2.Transf}. The terms $\sin(m\phi_0-\varphi_{m})$ and $\cos(m\phi_0-\varphi_m)$ in the right hand side of Eq.~\ref{Eq:WarpEv.vzVZ} can be rewritten as

\begin{gather}
\sin(m\phi_0-\varphi_m)=\sin(m\phi_0)\cos(\varphi_m)-\cos(m\phi_0)\sin(\varphi_m) \\
\cos(m\phi_0-\varphi_m)=\cos(m\phi_0)\cos(\varphi_m)+\sin(m\phi_0)\sin(\varphi_m). 
\end{gather}

Regrouping the terms as a function of $\phi_0$ and using the orthogonality of the Fourier modes, we obtain that the amplitudes $a^V_m,b^V_m$ from the $V_z$ fit are related to the amplitudes $A_m$ and $\varphi_m$ from the $Z$ and the warp evolutionary terms as 

\begin{gather} 
a^V_m = A_m[m\Omega-\dot{\varphi}_{m}(t_0)]\cos(\varphi_m)-\dot{A}_m(t_0)\sin(\varphi_m)\\
b^V_m = A_m[m\Omega-\dot{\varphi}_{m}(t_0)]\sin(\varphi_m)+\dot{A}_m(t_0)\cos(\varphi_m).
\end{gather}

Solving this linear system of equations for $\dot{A}_m(t_0)$ and $[m\dot{\phi}(t_0)-\dot{\varphi}_{m}(t_0)]$ and writing back  $a^V_m,b^V_m$ in terms of the amplitude and phase ($V_m,\varphi_m^V$), the evolutionary terms of the warp are given by

\begin{equation}
[m\Omega-\dot{\varphi}_{m}(t_0)]=\frac{V_m}{A_m}\sin(\varphi_m-\varphi_m^V)
\label{Eq:DiffAngularVel_n}
\end{equation}

and

\begin{equation}
 \dot{A}_m(t_0)=V_m\cos(\varphi_m-\varphi_m^V).
\label{Eq:Adot_n}
\end{equation}

Assuming that the $m$-th mode has angular velocity $\omega_m$, then setting $\varphi_m=m\omega_m t+\varphi_{m,0}$ in Eq.~\ref{Eq:DiffAngularVel_n}, we get the pattern speed for each mode as

\begin{equation}
\omega_m=\Omega-\frac{V_m}{mA_m}\sin(\varphi_m-\varphi_m^V).
\label{Eq:omega_n}
\end{equation}

Therefore, having connected the Fourier fits in $Z$ and $V_z$ at a given radius, Eqs.~\ref{Eq:Adot_n} and \ref{Eq:omega_n} describe how each pattern speed and amplitude change in time, allowing a reconstruct the time evolution of the warp as a function of radius.

We leave for a future work the publication of a more general framework that consider an azimuthal dependence not only of the vertical motion of the stars, but also their radial and azimuthal velocity, which would presumably result in a better inference of the time evolution of individual Fourier modes of the warp. \footnote{Using this framework with standard values allow us to conclude that the radial bulk motions and spiral arms can be ignored in a first-order approximation to derive the pattern speed and change in amplitude of the warp.}

\subsection{The Inference}\label{s:inference}

We have used Bayesian Inference to infer the $a_m,b_m$ (or $A_m$, $\phi_m$) that best describe the mean behaviour of the stars in a given ring when applying the methods from Secs.~\ref{sec:fourier_zvz} and \ref{Sec:AdotOmega} to a particular sample. Bayes' theorem \citep[e.g.][]{BayesSivia} relates the Posterior distribution to the Likelihood ($\mathcal{L}$)  and the Prior ($p$) probability densities functions (PDF) as

\begin{equation}
    P(\mathbf{X}|D,I) \propto \mathcal{L}(D|\mathbf{X},I)p(\mathbf{X}|I),
\end{equation}

\noindent where $D$ refers to the data, $\mathbf{X}$ to the model parameters and $I$ stands for any other available information. In our case, $\mathbf{X}$ is a vector containing the linear amplitudes: 


\begin{equation}
\mathbf{X}=[a_0,a_1,...,a_M,b_1,...,b_M].
\end{equation}
 
For the model parameters we assume uniform priors, with sufficiently and arbitrarily large limits. 
Assuming that the observations are independent, the likelihood is expressed as the product of the individual likelihood of each single data point $z_i$, for which we assume a Gaussian distribution

\begin{equation}
    \mathcal{L}(\{z\}|\mathbf{X},I)=\prod_{i=1}^N \frac{1}{\sqrt{2\pi\sigma_i^2}}\exp{\bigg[-\frac{(z_i-Z(\phi_i,\mathbf{X}))^2}{2\sigma_i^2}\bigg]}
    \label{eq:likelihood}
\end{equation}

\noindent where we take $\sigma_i^2$ to be the square sum of the uncertainty in the measurement $z_i$ and the \emph{intrinsic dispersion} $\sigma_{ID}$ of the variable, at that ring. This $\sigma_{ID}$ is introduced to take into account the natural dispersion around the mean value that arises from the dynamics of the Galactic disc; in $v_z$, it measures the velocity dispersion and, in $z$, it measures how thick the disc is at that ring. The intrinsic dispersion is not a free parameter in the fit. We estimate it as the mean dispersion in the variable of interest in equally spaced azimuthal bins, weighted by the number of stars in each bin because low number statistics dominate over observational errors. 

In our case, because the model is linear in all parameters and we have assumed a uniform prior, the maximum a posteriori (MAP) $\mathbf{X_0}$ coincides with the maximum of $\mathcal{L}$ and the posterior is exactly a Gaussian distribution with mean $\mathbf{X_0}$ and covariance matrix $\mathbf{\Sigma}$ \citep[see e.g. Sec.~1 in][for a detailed discussion]{Hogg2010}. The posterior PDF can, therefore, be expressed as

\begin{equation}
    P(\mathbf{X}|D,I)=\frac{\exp\big[-\frac{1}{2}(\mathbf{X}-\mathbf{X_0})^T\mathbf{\Sigma}^{-1}(\mathbf{X}-\mathbf{X_0}) \big]}{(2\pi)^{N+1/2}\big[\det{\mathbf{\Sigma}}\big]^{1/2}}
    \label{eq:posterior}
\end{equation}

\noindent where $\mathbf{X_0}$ is given by

\begin{equation}
   \mathbf{A}\mathbf{X_0}=\mathbf{p}. 
\end{equation}

The covariance matrix $\mathbf{\Sigma}$ is the inverse of $\mathbf{A}$: the matrix that contains in its entries the "projection" of each mode into the other ones (see Eq.~\ref{Eq:App.matrixA} in Appendix~\ref{app:Post_Sol}) weighted by the dispersion in the data, and the vector $\mathbf{p}$ has the "projection" of the data in each mode (see Eq.~\ref{Eq:App.VecP}). 

Because we use a Fourier series to represent a variable, one would expect the modes to be mutually independent and therefore not correlated. This is not usually the case. When we have discrete measurements, the modes are not mutually orthogonal unless the measurements are equally spaced in azimuth and have the same $\sigma_i$. In this special case $\mathbf{A}$ is diagonal and, in consequence, the covariance matrix is too. This particular distribution allows the modes to be mutually independent. Naturally, we will never get this configuration from the data itself, but this method shows analytically that data that are more or less uniformly distributed in azimuth are preferred for a Fourier analysis of the whole disc: studies with a sparse and irregular azimuthal coverage will get modes that are not "fundamental", in the sense that \emph{they are not describing the modes of the warp itself}. The effect gets worse with high-frequency modes $m\geq2$. This should be kept in mind when interpreting results for individual modes, nevertheless, it will not affect our conclusions on the description of the warp as a whole (the sum of the modes) in the regions well sampled by the data. 


Finally, the disc is divided in rings such that we get a "continuum" view of how the modes and the warp change with the radius. To do so we take each ring to contain a fixed number $N$ of stars out of the total $N_\mathrm{tot}$ stars, the first ring starting with the star at the smallest radius. The second ring will start at the radius of the second star and have a width such that it also contains $N$ stars, and so on for subsequent rings. This scheme implies that the rings will have a varying width, depending on the sample's radial distribution. We take the radius associated with each ring as the mean radius of the stars in it. This procedure allows us to have a continuous view, with all rings having the same number of stars $N$ and, therefore, constant stochastic noise. It must be kept in mind, however, that only one out of every $N$ consecutive rings will be independent. 
Changing the number of stars in each ring changes the smoothened parameters inferred as a function of the radius (the bigger N, the smoother it gets). Also, the change in N moves the mean radius of each ring, the tendency is that a bigger N makes the rings to move inwards (smaller radius), as expected for a density profile that decreases with radius.

\subsubsection{Validation with simulations}\label{sec:validation_sims_summary}

Here we present our main conclusions about the performance of the methods described in the previous section, assessed by applying them to mock catalogues constructed from test particle simulations. As discussed in detail in Appendix.~\ref{sec:FDM}, we used a test particle simulation of a warped disc from \citet{RG19} to create a mock catalogue affected by the Gaia DR3 selection function (SF) and observational errors. A fiducial model, unaffected by the SF or by errors, is used as a baseline for comparison of the results of the Fourier Decomposition. The interested reader may find full details and discussion of these results in Appendix.~\ref{sec:FDM}. 

Our main results on how the SF affects the recovery of the warp as a whole, in different regions of the disc, are summarised here as follows:

\begin{itemize}
    \item For $Z$ the best sampled region, the quadrants I and III ($X<0$~kpc) is recovered well (differences between the real and the recovered warp are smaller than $\sigma_{ID}$) and the general tendency for all radii is recovered for both series summing up to $M=1$ and $M=2$. For $X>0$~kpc (quadrants I and IV) the SF causes the warp to be exaggerated. This bias is reduced for outer radii as the main mode of the warp ($m=1$) becomes greater than the intrinsic dispersion (see Fig.~\ref{fig:DeltaModel}).
    \item For $V_z$ the recovery is better than $Z$, although for the inner disc ($R\lessapprox9$~kpc) the recovery is poor for $X>0$. The recovery in the sampled area is better than in $Z$ for both $M=1$ and $M=2$ (differences are smaller than $\sigma_{ID}^V$ in most of the disc area, see Fig.~\ref{fig:DeltaModel}).
    \item Main conclusion: \emph{The recovery of the full model (the Fourier sum) in both variables is robust in the well sampled regions for $R>10$~kpc, i.e. second and third quadrants. In this region all the warp features are well recovered within the uncertainties given by the Posterior realisations. } 
\end{itemize}

We also tested how the individual modes are recovered, our main conclusion being: 

\begin{itemize}
    \item The $m=0$ mode is well recovered throughout the disc.
    \item The $m=1$ mode tends to be overestimated in amplitude at the inner disc.In the outer regions ($R>10$~kpc) its amplitude and phase are well recovered in both variables (i.e $Z$ and $V_z$).
    \item The $m=2$ mode captures the asymmetries and is well recovered where there is data, i.e. quadrants II and III and $R>10$~kpc, but tends to be underestimated in amplitude and the general trend of the phase is poorly recovered in the whole disc.
    \item Main conclusion:\emph{ The uncertainty on the recovery of the individual modes stems from the correlations between the modes which, in turn, appear as a consequence of the imperfect azimuthal coverage. The degeneracies introduced by the correlations mean that different combinations of amplitudes and phases for the individual modes can give the same sum model, in a finite azimuth range. A full azimuthal coverage would break this degeneracy and make the inference on the individual modes unique. The mode less affected by this degeneracies is the $m=1$ mode due to its large amplitude.}
\end{itemize}

The intrinsic dispersion in $z$ is well recovered in the outer disc where the warp amplitude is larger than the dispersion. In the inner disc $\sigma_{ID}$ tends to be off by $10\%$. For $v_z$, we find $\sigma_{ID}^V$ is underestimated by $3\%$ without dependency on the radius. 

Given these results, we decide to include up to the $m=2$ mode in the Fourier fits for this work because it offers the least biased recovery for the region of the disc where the warp is most prominent (i.e. outer radii). Reliable results for the inner region of the disc are limited to $|\phi|\gtrsim90^\circ$, the region least affected by the SF with best coverage, where biases in the recovery are lowest.\\

We also tested the inference of the time evolution parameters $\dot{A}_m$ and $\omega_m$. We concluded that the recovery of the $\dot{A}_m$ for $m=2$ is unreliable due to the biases and noise. For the $m=1$ mode we conclude:

\begin{itemize}
    \item $\dot{A}_1$ is well recovered within its uncertainties particularly for the outer disc 
    \item The recovered $\omega_1$ tends to be overestimated due to a slight overestimation in $A_1$, but the mean difference is $\approx4$~\kms kpc$^{-1}$. In the outer disc ($R>14$~kpc) we recover the values of the fiducial model within the uncertainty.
\end{itemize}

\section{The Cepheids Sample}\label{sec:Sample}

 We use the catalogue of Milky Way Cepheids from \cite{Skowron2020}. The catalogue contains $2385$ Classical (Type I) Cepheids identified mainly with the OGLE survey \citep[for more details see][]{Skowron2019, Skowron2020} with photometric distances computed based on mid-IR photometry from the Wide-field Infrared Survey Explorer (WISE) and the Spitzer Space Telescope, resulting in distance uncertainties of $3\%$ on average. We cross-matched the Cepheid catalogue (at $1$~arcsec tolerance) with Gaia DR3 to retrieve proper motions for these stars. Out of the $2381$ Cepheids with Gaia proper motions, only 860 stars have radial velocities in DR3. In order to curate a homogeneous catalogue with full velocity information allowing us to compute $v_z$, we infer the missing line of sight velocity for all stars in the catalogue by assuming the Cepheid rotation curve from \citet{V(R)_Ablimit} which has a slope of $-1.33$~\kms$\text{kpc}^{-1}$ and takes the value $232$~\kms\ at the solar radius\footnote{We have tested different values for the rotation curve around this one and our results are not affected.}.

We clean this sample by keeping stars with RUWE~$<1.4$, $\sigma_z\leq0.1$~kpc and $\sigma_{v_z}\leq13$~\kms. These upper bounds in $z$ and $v_z$ uncertainties guarantee a significant amount of stars whose uncertainties are at most of the order of $\sigma_{ID}$. To avoid clear outliers due to probable contaminants and the Magellanic Clouds we restrict the analysis to stars with $|z|\leq2$~kpc, $|v_z|\leq30$~\kms and $3~\text{kpc}<R<18$~kpc. These are very broad cuts that only remove very few ($\approx3\%$) clear outliers (5 sigmas) most of them due to the cut in $V_z$ (only one star is removed for the cut in $Z$). These constraints reduce the sample to a total of $N_\mathrm{tot}=1997$ stars.

\section{Results}\label{sec:ResultsFULL}

Here we present the results obtained by applying the methods described in Sec.~\ref{sec:FDM} to the final sample with $M=2$ and $N=200$ stars in each ring. To calculate $\sigma_{ID}$ in both variables we use $8$ azimuth bins. The resulting amplitudes and phases as a function of radius for the best fitting (Maximum a Posteriori, MAP) models for $Z$ and $V_z$ are provided in Table~\ref{tab:results} and 100 posterior realisations are provided in Table~\ref{tab:result_posteriors}. Figures \ref{fig:A_R} and \ref{fig:Phi_R} in the Appendix~\ref{app:ResultsIndividualModes} show the amplitudes and phase (respectively) of each mode in $Z$ and in $V_z$ as a function of the radius. 

\begin{table*}
    \centering
	\caption{Amplitudes and phases as a function of radius for the best fitting (MAP) models for $Z$ and $V_Z$. The first few rows of the table are shown to provide guidance regarding its form and content. The full version of the table is available in the electronic version.}
	\label{tab:results}
	\begin{tabular}{lcccccccccr} 
		\hline
          $R$ & $A_0\sin(-\varphi_0)$ & $A_1$ & $A_2$ & $\varphi_1$ & $\varphi_2$ & $V_0\sin(-\varphi_0^V)$ & $V_1$ & $V_2$ & $\varphi^V_1$ & $\varphi^V_2$ \\
          (kpc)& (kpc)& (kpc)& (kpc)& (rad)& (rad) & (\kms)& (\kms)& (\kms)& (rad)& (rad)\\
		\hline
		   5.171&   -0.106 &  0.171 &  0.059 &  1.971705&    2.337575  & -0.422 &  2.638  & 1.976 &  -2.887806 &  -2.663989    
 \\
            5.186 &  -0.107  & 0.176  & 0.063 &  1.979711    &2.40411  &  -0.327 &  2.455  & 1.865   &-2.875007 &  -2.720106 \\
            5.202  & -0.107 &  0.175  & 0.062 &  1.978782&    2.403354  & -0.363  & 2.296  & 1.941&   -2.777056 &  -2.595807\\  
		\hline
	\end{tabular}
\end{table*}

\begin{table*}
    \centering
	\caption{Amplitudes and phases as a function of radius for 100 posterior realisations for the $Z$ and $V_Z$ models. The first few rows of the table are shown to provide guidance regarding its form and content. The full version of the table is available in the electronic version.}
	\label{tab:result_posteriors}
	\begin{tabular}{lcccccccccr} 
		\hline
          $R$ & $A_0\sin(-\varphi_0)$ & $A_1$ & $A_2$ & $\varphi_1$ & $\varphi_2$ & $V_0\sin(-\varphi_0^V)$ & $V_1$ & $V_2$ & $\varphi^V_1$ & $\varphi^V_2$ \\
          (kpc)& (kpc)& (kpc)& (kpc)& (rad)& (rad) & (\kms)& (\kms)& (\kms)& (rad)& (rad)\\
		\hline
		  5.171  & -0.109 &  0.163 &  0.042  & 2.000795&   2.192443 &  -0.951   &3.761 &  3.294 &  2.232128&    1.993639 \\
             5.171  & -0.097  & 0.155&   0.053  & 1.848048&   2.048916&   0.626 &   5.258  & 4.114   &-2.853366  & -3.10008\\
            5.171  & -0.111  & 0.19  &  0.076 &  1.897841&   2.256799   &-3.56 &   3.122  & 1.602  & 2.53726   &  -2.602042\\ 
		\hline
	\end{tabular}
\end{table*}

\subsection{Structure and kinematics of the warp}
\label{sec:Results}

In the following sections we analyse different features of the warp structure and kinematics. We  analyse the full Fourier series obtained. Since the validation with simulations indicated results for the individual modes are prone to be biased due to correlations between the modes, we discuss and summarise this in Appendix~\ref{app:ResultsIndividualModes} for the interested reader. 

\subsubsection{General structure of the warp}

\begin{figure*}
	\includegraphics[width=17cm]{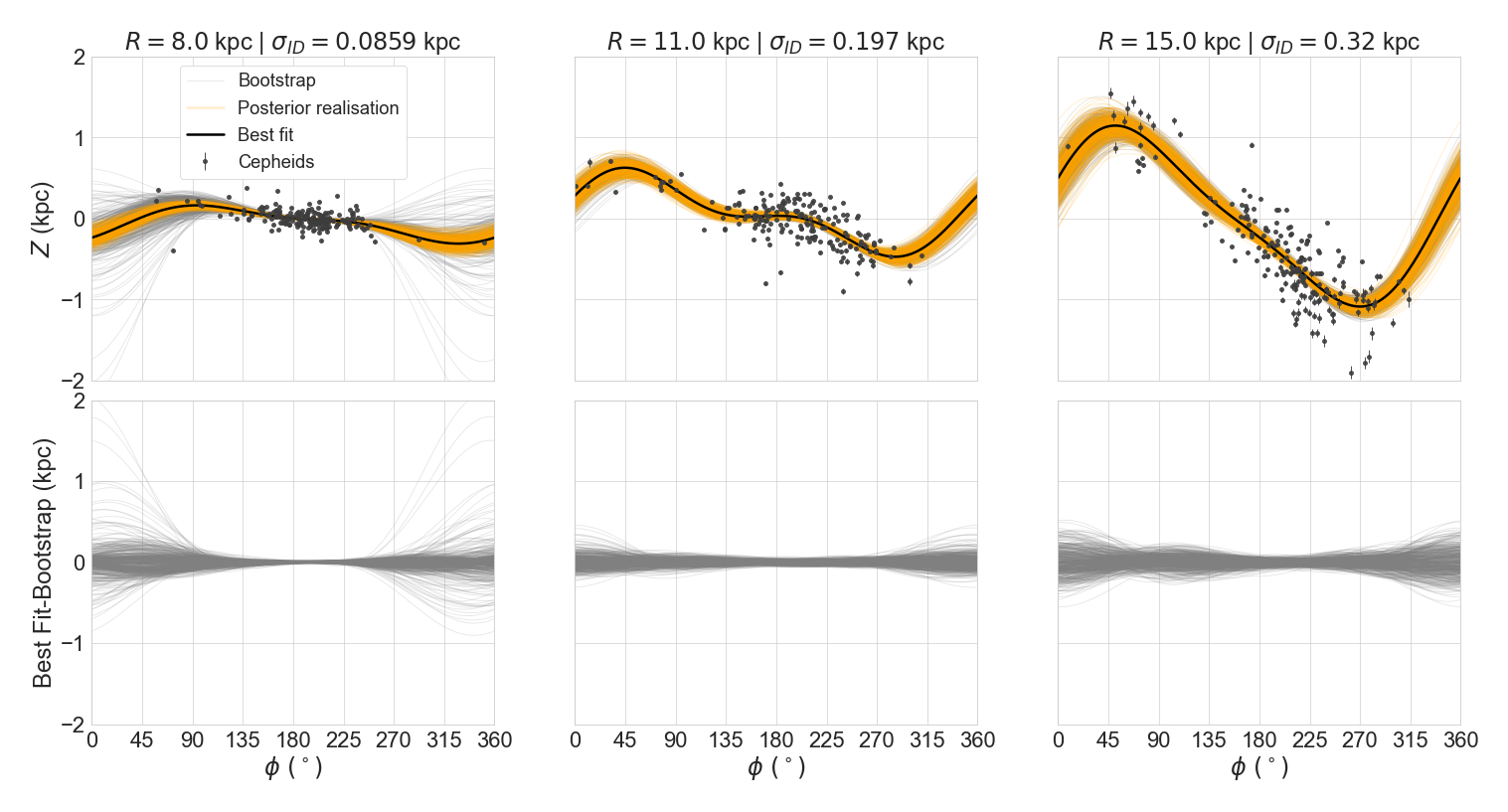}
    \caption{The upper panels are the vertical height $Z$ as a function of galactocentric azimuth $\phi$ for three different rings with radius $8.0$~kpc (left), $11.0$~kpc (middle) and $15$~kpc (right). Gray dots represent the Cepheids in each ring, the black solid line represents the Fourier curve MAP fit to the gray dots, and the oranges curves are $500$ random realisations of the Posterior PDF. The gray curves are fits to $200$ bootstrapping realisations. The bottom panels show the residuals between the best fit and each bootstrap realisation.}
    \label{fig:Z_phi_3panels}
\end{figure*}

We show in the upper panels of Fig.~\ref{fig:Z_phi_3panels} the results of three fits in $Z$ for different Galactocentric radii. Each panel shows, for rings of increasing radius, the Cepheids present in the ring, the best  Fourier fit (black curve) and 500 random realisations from the Posterior PDF, the gray curves are fits to $200$ bootstrapping realisation. The plots clearly show a growth in amplitude typical of an S-shaped warp, reaching a maximum of $\approx 1.1$~kpc in the outskirts of the disc. The effect of the SF is evident, the azimuth range sampled increases with radius. Other features like the change of the warp as a function of $\phi$ become clear in the second panel ($R=11.0$~kpc), where a plateau is noticeable around $\phi=180^\circ$. The third panel ($R=15$~kpc) shows how from $\phi\approx60^\circ$ to $\phi\approx 240^\circ$ the change in the warp between the extremes resembles a straight line more than a sinusoidal curve corresponding to a pure $m=1$ mode would. This feature is correctly reproduced by the model thanks to the $m=2$ mode; a simple tilted rings model ($M=1$) cannot reflect it. The fits to the bootstrapping realisations shows that for $R<10$~kpc the fits are affected by statistical noise as shown in the first panel of Fig.~\ref{fig:Z_phi_3panels} at $R=8$~kpc in the first and fourth quadrant of the galactic plane, this became more clear in the residuals plot in the bottom panels. For the outer radii the fits are less sensitive to statistical noise as we see in the second and third panel where the posterior realisations coincide with the bootstrap realisations. For this reason we focus our \Mauro{analysis} on the second a third quadrants. 

\begin{figure*}
	\includegraphics[width=17cm]{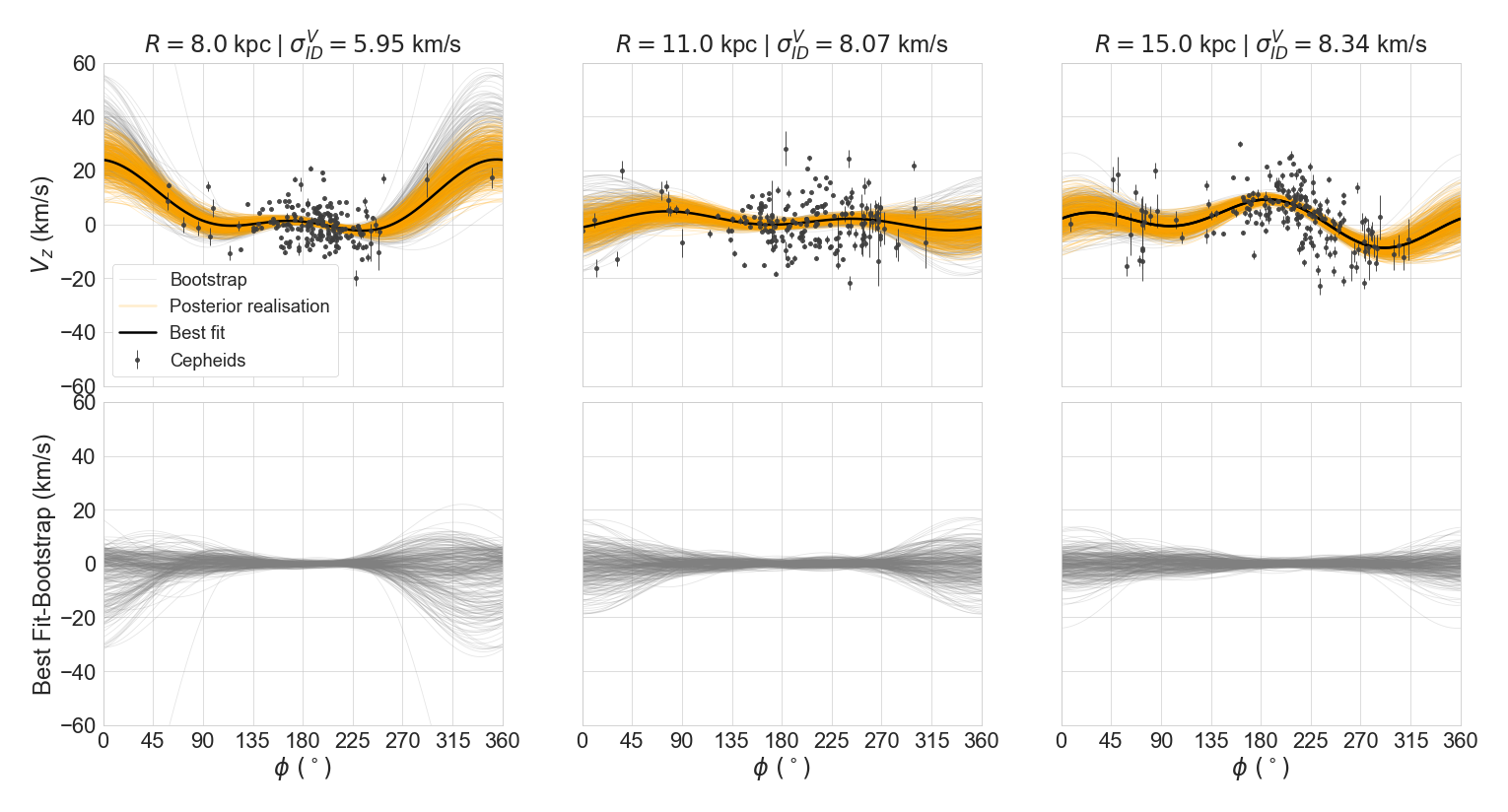}
    \caption{The upper panels are the vertical velocity $V_z$ as a function of galactocentric azimuth $\phi$ for three different rings with radius $8.0$~kpc (left), $11.0$~kpc (middle) and $15$~kpc (right). Gray dots represent the Cepheids in each ring, the black solid line represents the Fourier curve MAP fit to the gray dots, and the oranges curves are $500$ random realisations of the Posterior PDF. The gray curves are fits to $200$ bootstrapping realisation. The bottom panels show the residuals between the best fit and each bootstrap realisation.}
    \label{fig:Vz_phi_3panels}
\end{figure*}

Figure~\ref{fig:Vz_phi_3panels} shows $V_z$ as a function of $\phi$ for the same three rings shown in Fig.~\ref{fig:Z_phi_3panels}. The first panel ($R=8.0$~kpc) of this figure, as well as in the previous one, shows how the few observed data points in regions most affected by the SF (e.g. $\phi \sim 300^\circ$ ) strongly drive the fit in those regions. As discussed in Sec.~\ref{app:ResultsIndividualModes}, this makes the inference unreliable for the inner disc at $R<10$~kpc, except around the azimuth of the solar neighbourhood. Therefore, in what follows we will restrict our analysis to $R\geq10$~kpc. As radius increases (second and third panels) the amplitude of the warp in velocity grows but only mildly, as it is at most of the order of the intrinsic dispersion $\sigma^V_{ID}\approx8$~\kms\ even at the outer disc. This is in contrast with $Z$, where the amplitude of the warp exceeds the intrinsic dispersion by a factor of $\approx 3$ in the outer disc. This low amplitude in comparison with $\sigma^V_{ID}$ makes it harder to detect the kinematic signature of the warp, but at the outer disc it is clear there is a complex and asymmetrical behaviour, as seen in the third panel in Fig.~\ref{fig:Vz_phi_3panels}. The bootstrap realisations for $V_z$ show the same conclusions as in $Z$, but due to the low amplitude of the kinematic signal in the first and fourth quadrant the realisations show a greater dispersion than the posterior, illustrating that due to low number statistics noise is larger. For this reason we will focus the analysis of the kinematic signal to the second and third quadrants.

\subsubsection{Asymmetries in height }
\label{sec:Asymmetries}

\begin{figure*}
	\includegraphics[width=17cm]{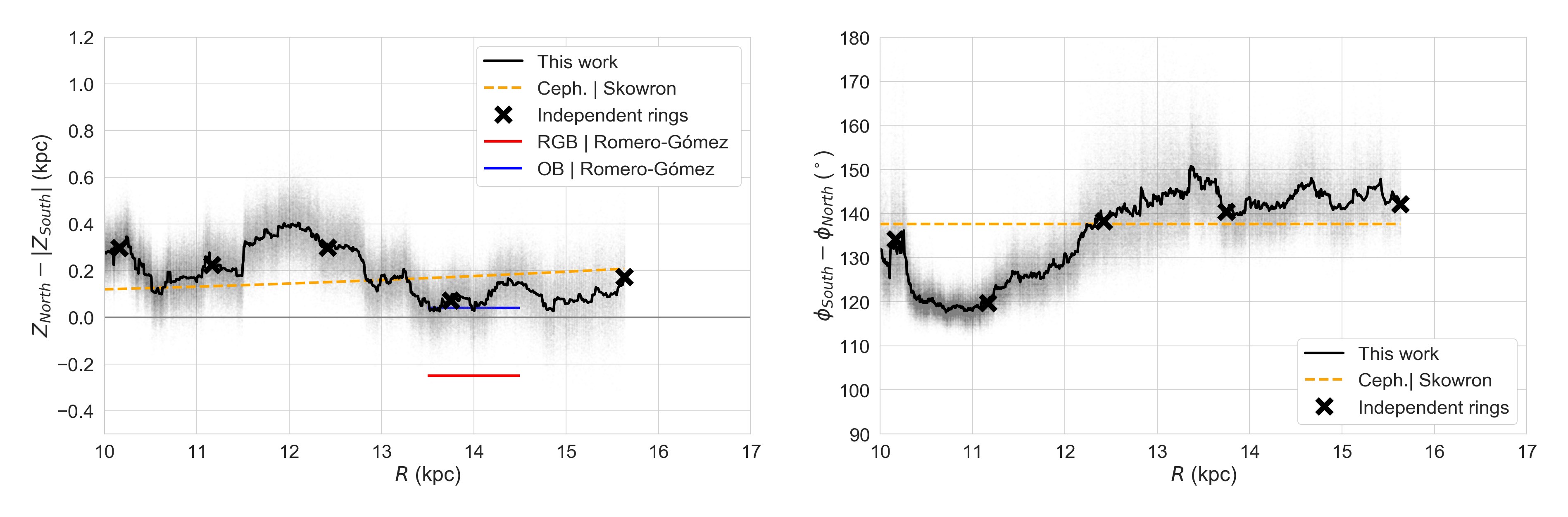}
    \caption{Left: Difference between the north and south extreme of the warp as a function of galactocentric radii from our results (black curve), the same is calculated for the warp model by \citet{Skowron2020} (doted orange curve). Right: Least angular difference between the north and south extremes as a function of galactocentric radii from our results (black curve), the same is calculated for the warp model of \citet{Skowron2020} (doted orange curve). The grey dots are $500$ random realisation at each ring taken from the Posterior.}
    \label{fig:Asym}
\end{figure*}

First, we explore the asymmetries of the warp in height above and below the plane. The left panel of Figure~\ref{fig:Asym} shows the difference between the maximum and minimum height reached by the warp above and below the plane in the North and South Galactic hemispheres respectively. Positive values in this plot, at any given radius, imply that the Northern extreme of the warp deviates more from the Galactic plane than the South. Up to $R\approx12$~kpc the northern extreme is larger than the southern, even within the uncertainties, showing an asymmetrical warp. This asymmetry decreases towards the outer disc, with the warp being almost symmetrical to within the uncertainties ($\approx100$~pc) at $R\gtrsim13.5$kpc. We should keep in mind that because of the SF, the extremes of the warp tend to be overestimated in the internal regions. However, a more accurate and reliable measurement of this asymmetry is expected at the outskirts of the disc from our validation tests (Sec.~\ref{sec:validation_sims_summary}). 

Since we set the phases of each mode free, we can also track the azimuth of each extreme of the warp to explore the \emph{azimuthal} asymmetry as a function of $R$. The right panel of Fig.~\ref{fig:Asym} explores the azimuthal asymmetry of the extremes of the warp as a function of radius by showing the smallest angular difference in the azimuths of the warp extremes in $Z$. In a simple tilted rings model of an S-shaped warp these extremes are always separated $180^\circ$, even if the line of nodes is twisted.
The plot clearly shows \emph{the extremes of the Cepheid warp are never diametrically opposed}. The difference in azimuth starts at its lowest value of $\approx120^\circ$ at $R\approx10-11.5$~kpc and increases up to $\approx145^\circ$ at $R\approx12.5$~kpc after which it remains approximately constant. This is a robust feature that cannot be reproduced by an $m=1$ warp, reinforcing the need for an $m=2$ mode to describe the full warp. 

\subsubsection{Line of nodes and Line of Maximum $V_z$}
\label{sec:LON_LMV}

\begin{figure}
	\includegraphics[width=9cm]{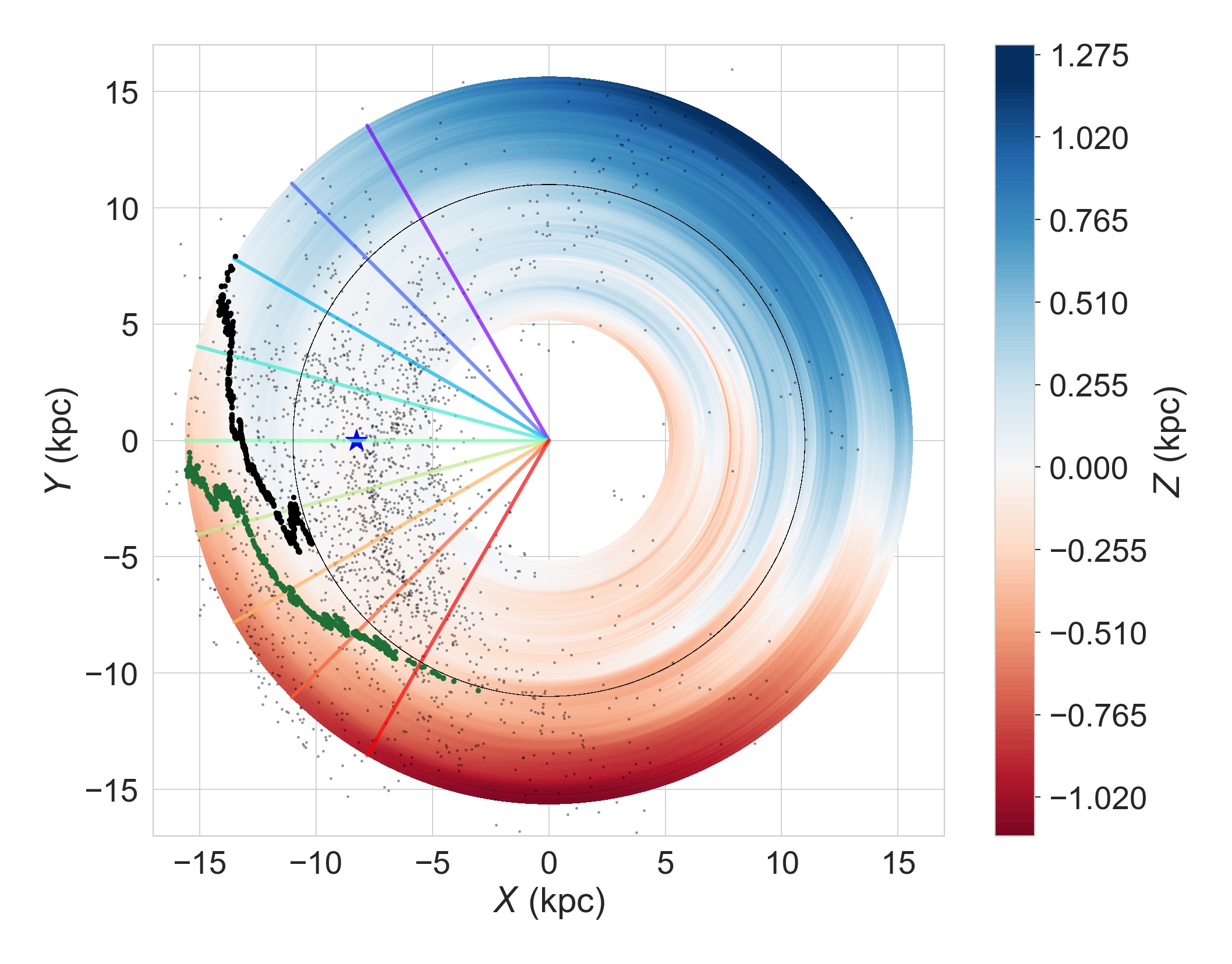}
    \caption{Face-on view of the best fitting (MAP) warp model for the Cepheids (gray dots). The colour scale represents the mean Z of the disc (blue above the plane and red below it). The line of nodes (LON, i.e. $Z = 0$) is indicated with the black curve. The line of maximum vertical velocity (\LMV) is indicated by the dark green curve. The different coloured lines correspond to lines of constant galactocentric azimuth.}
    \label{fig:XYlon}
\end{figure}

The overall behaviour of the best fitting (MAP) warp model for the Cepheids is shown in Fig.~\ref{fig:XYlon} in a face-on view of the disc with a colour scale indicating the mean height above/below the mid-plane. The line of nodes (from now on LON) and line of maximum vertical velocity (\LMV) are indicated with the black and green lines respectively. A leading twist (i.e. in the direction of Galactic rotation) in both the LON and \LMV\ is evident, as well as an offset between the two. 

\begin{figure}
	\includegraphics[width=9cm]{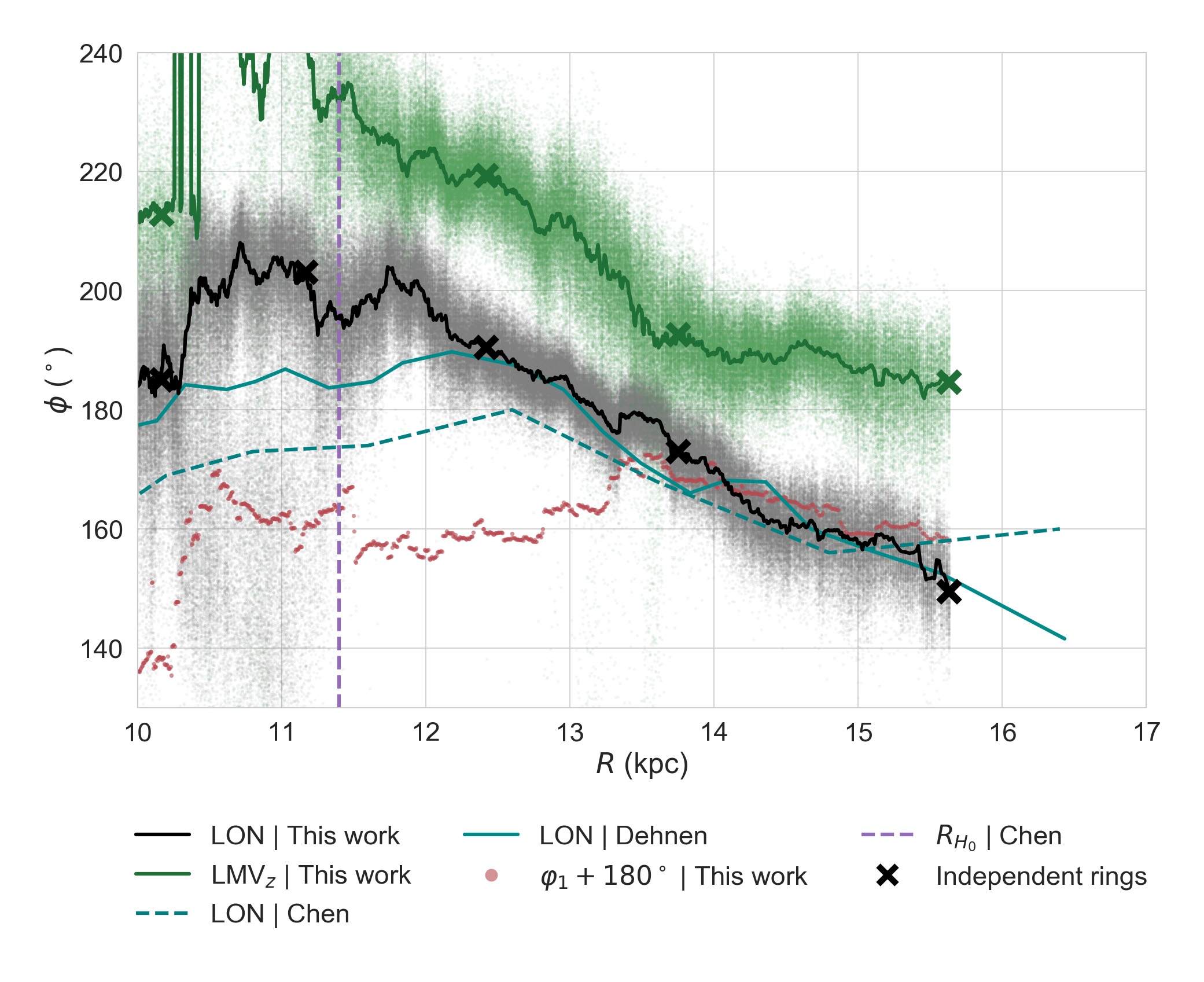}
    \caption{Azimuth as a function of galactocentric radius for: the LON from this work (black curve), \citet{Chen3Dmap} (cyan dashed curve) and \citet{Dehnen_2023_Warp_Ceph} (red curve); the \LMV\ (thick green curve) and $\varphi_1$ for $m=1$ mode from our $Z$ fits (red dots). The gray and olive green dots are $500$ realisations of the LON and the \LMV\ taken from the posterior at each ring, respectively. The vertical dashed line indicates the Holmberg radius for the Milky Way from \citet{Chen3Dmap}. The crosses indicate a sample of independent (disjoint) rings.}
    \label{fig:LON}
\end{figure}

Fig.~\ref{fig:LON} shows the LON and \LMV\ azimuths (for $X<0$) as a function of radius. The figure shows that the azimuth of the LON is well represented by the straight line (in the plane $\phi,R$) with the parameters presented in Eq.~\ref{Eq:3LONfit}, obtained from a fit to data in independent rings with $R>11$.  
The \LMV\ also follows a linear tendency, well described by an almost constant azimuthal difference of $25.4^\circ$ with respect to the LON. 

\begin{equation}
    \phi_{\text{LON}}(R)=(-12.7\pm0.3)\dfrac{\text{deg}}{\text{kpc}}R+(347\fdg5\pm3\fdg5)  \ \ \ \ \text{for} \ \ \ \ R>11~\text{kpc}.
    \label{Eq:3LONfit}
\end{equation}

\subsection{The time evolution of the warp traced by Cepheid}
\label{Sec:TimeEvol}

Here we present results for the pattern speed (from Eq.~\ref{Eq:omega_n}) and the change in amplitude with time (from Eq.~\ref{Eq:Adot_n}) for the $m=1$ mode obtained for the Cepheids. 
We ignore the $m=0$ mode, since its pattern speed is ill-defined and its amplitude change is $V_0\sin(-\varphi_0^V)$ (this is shown in the right panel of Fig.\ref{fig:A_R}). 

Although the Fourier series for $Z$ and $V_z$ have been fit with $M=2$, we focus this analysis in the $m=1$ mode,  because it is the dominant mode of the warp and the recovery of the evolutionary terms for $m=2$ are biased and noisy due to selection function effects (as shown in Sec.~\ref{sec:RecOfModes}). From Sec.~\ref{sec:validation_sims_summary} we recall that, for our simulation, $\omega_1$ and $\dot{A}_1$ are well-defined for $R\gtrsim12$~kpc where $V_1$ is non-zero and also well-defined (as shown in Sec.~\ref{app:ResultsIndividualModes}). Therefore, we will restrict this part of the analysis to $R\gtrsim12$~kpc. 
 
\subsubsection{Amplitudes}

\begin{figure}
	\includegraphics[width=\columnwidth]{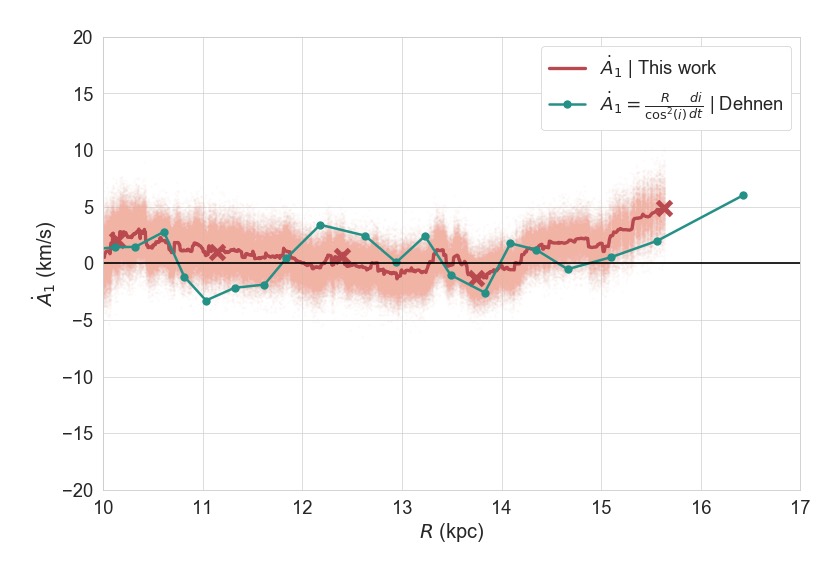}
    \caption{The change in amplitude $\dot{A}_1$ for the $m=1$ mode as a function of the galactocentric radii from our fits (black curve) and the $\dot{A}_1$ by \citet{Dehnen_2023_Warp_Ceph} (cyan curve). Red dots indicate measures for independent rings. The gray dots around each $\dot{A}_1$ are $500$ realisation taken from the posterior at each ring.}
    \label{fig:Adot}
\end{figure}

In Fig.~\ref{fig:Adot} we present results for $\dot{A}_1$ as a function of $R$. In the range $R<14.5$~kpc the change in amplitude is negligible, for $R>15$~kpc it shows a growing tendency\footnote{We consider this result to be extended beyond $R=15$~kpc and not only up to $R=15.5$~kpc because the rings at this radius contain stars beyond $R=15.5$~kpc} reaching a maximum in the external disc of $\approx 5$~\kms$\approx5\text{kpc Gyr}^{-1}$, this tendency is also present in the results by \citet{Dehnen_2023_Warp_Ceph}. Based on our validation summarised in Sec.~\ref{sec:validation_sims_summary} we expect these results to be unbiased over this radial range.

\subsubsection{Pattern speed}

\begin{figure}
	\includegraphics[width=\columnwidth]{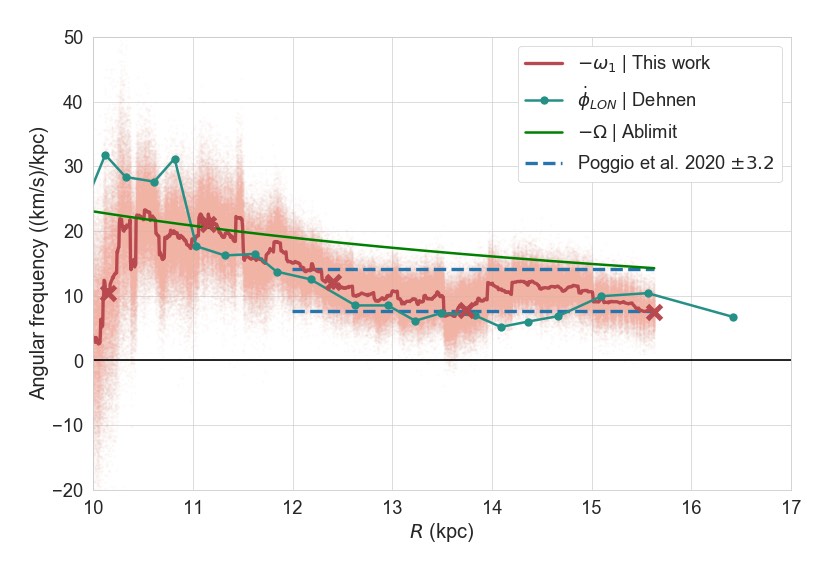}
    \caption{Minus the angular frequency as a function of the galactocentric radius for the pattern speed for the $m=1$ mode $\omega_1$ (black curve) from our fits, the angular velocity given by the rotation curve $\Omega$ (red curve) \citep{V(R)_Ablimit}, the upper and lower limit found by \citet{Poggio_Warp_evolving}, and results from  \citet{Dehnen_2023_Warp_Ceph}. The dots around $\omega_1$ are $500$ posterior realisations  at each ring.}
    \label{fig:Omega_1}
\end{figure}

Assuming the angular velocity $\Omega$ from the rotation curve by \cite{V(R)_Ablimit}, we  obtained the pattern speed for the $m=1$ mode from Eq.~\ref{Eq:omega_n}. Because in our reference frame the stars rotate in the direction in which $\phi$ decreases, the angular velocities in the direction of stellar rotation are negative. \emph{To avoid confusion we present the angular velocities with their sign changed}. Figure~\ref{fig:Omega_1} presents, as a function of Galactocentric radius, the pattern speed of the $m=1$ mode $\omega_1$ (black curve), the angular velocity of the rotation curve $\Omega$ (red curve), the upper and lower limit given by the measurements by \cite{Poggio_Warp_evolving} (dotted blue lines) and results from \citet{Dehnen_2023_Warp_Ceph} (solid cyan line and dots). 

We find that $\omega_1$ decreases for $R\gtrsim11$~kpc and shows a small oscillation for $13<R/\text{kpc}<16$. 
This overall behaviour, both the decrease and the oscillation, are observed by \citet{Dehnen_2023_Warp_Ceph} but at a slightly different radius. This difference may arise from their use of guiding radius and also because we use a mean radius to represent each ring, which tends to drive the results from the outer to the inner radii. We would need smaller uncertainties to ensure this oscillation is a physical phenomenon in the disc and not an artefact from our fits. However, the fact that it is also observed by \citet{Dehnen_2023_Warp_Ceph}, with a sample that includes radial velocities, increases our confidence in the result. Our mean value observed for $R>12$~kpc is in agreement with the results from previous works on measuring the pattern speed by \citet{Poggio_Warp_evolving} and \citet{Cheng_2020_Warp_patern}, who assumed rigid body rotation for the warp.

\section{Discussion}
\label{sec:Discussion}

\subsection{Structure and kinematics}

\subsubsection{Comparison with different warp observations}
\label{sec:Discussion_Comp} 

\begin{figure*}
	\includegraphics[width=16cm]{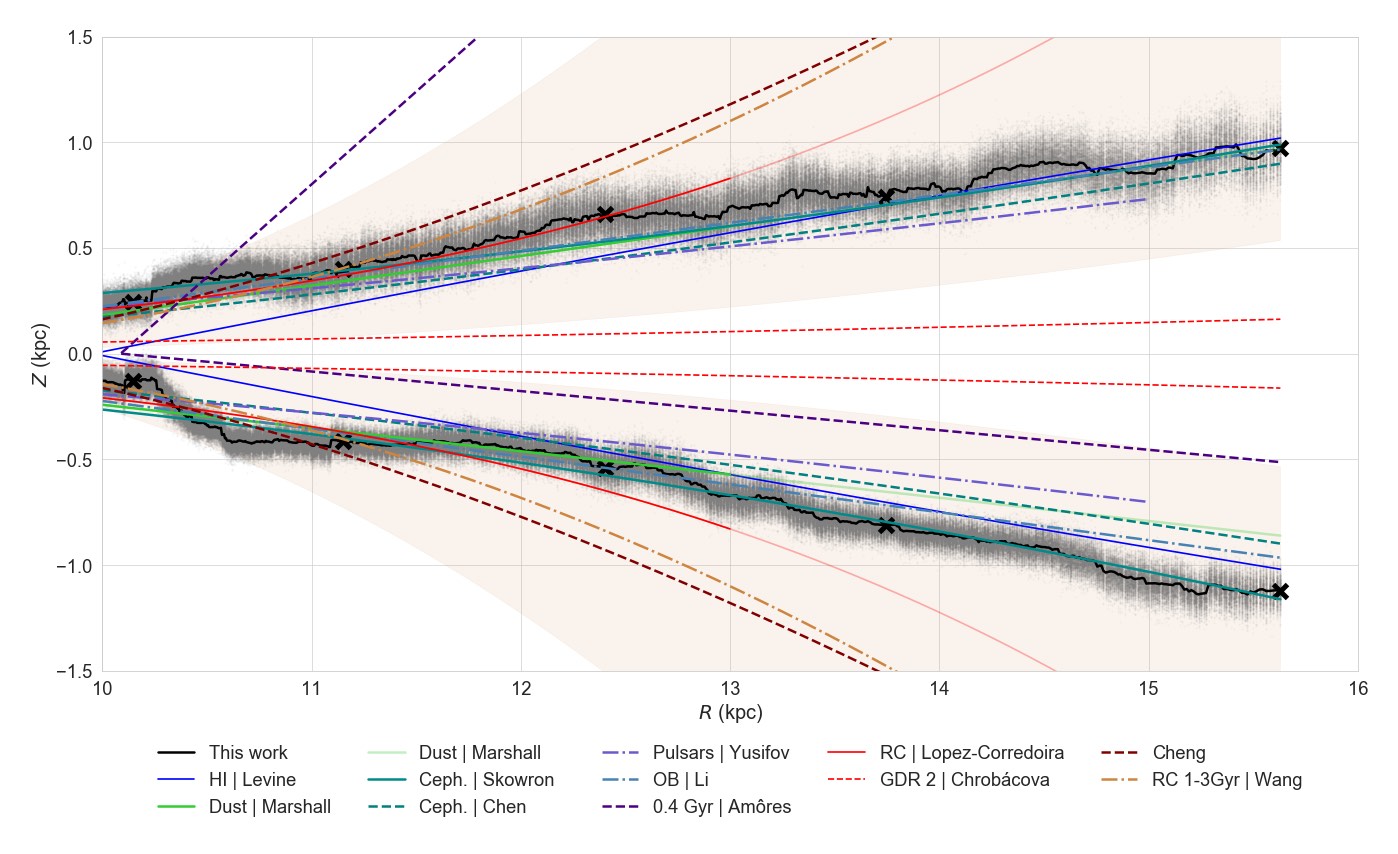}
    \caption{Vertical height of the warp as a function of galactocentric radius for slices at $\phi=90^\circ$ and $\phi=270^\circ$, for this work and warp models in the literature summarised in Table~\ref{tab:Models_Lit}. The shaded region represents the uncertainty in the warp model from \citet{Wang_LAMOST_Disk_Warp_RC_2020} (see text for more details).}
    \label{fig:NScomparison}
\end{figure*}

In Fig.~\ref{fig:NScomparison} we compare different warp models in the literature to our results for $\phi=90^\circ$ (northern region) and $\phi=270^\circ$ (southern region) for $R>10$~kpc. The various works cited here have different azimuthal and/or radial coverage, use different tracers, and have used different methods to fit for the warp. Table~\ref{tab:Models_Lit} summarises this information for the works presented in the figure. We have selected these works in order to compare against other dynamically young tracers like the gas, dust and OB stars. We also include results from a few warp models for dynamically older populations for which the time evolution of the warp has been inferred. Since the warp followed by the older population may differ from that of the young, in Sec.~\ref{sec:Disc_TimeEvol} we will discuss the effect due to the assumed structure on the inference of the time evolution of the warp.

\begin{table*}
	\centering
	\caption{Models from the literature. The asterisk ($*$) indicates the model shown in Fig.~\ref{fig:NScomparison}.}
	\label{tab:Models_Lit}
	\begin{tabular}{lcccr} 
		\hline
		Work & Tracer/Method & Symmetric (yes or no) & LON (fixed or free) & Disc range\\
		\hline
        \citet{Warp_Pulsars_Yusifov_2004} & Pulsars & Yes & Fixed & $R\lessapprox15$~kpc\\
        \citet{Levine2006} & HI & No & Fixed$^*$/Free & $10<R/\text{kpc}<30$\\  
        \citet{Marshall_Modelling_interstellar_extinction_2006} & Dust & No & Fixed & $[d\lessapprox13]$~kpc $l \in [-90^\circ, 90^\circ]$\\
		\citet{Skowron2020} & Cepheids & No & Fixed & $R<20$~kpc\\
		\citet{Chen3Dmap} & Cepheids & Yes$^*$/No & Fixed$^*$/Free & $5\lessapprox R/\text{kpc}<20$\\
        \citet{Li_2023_Nonsteady_Warp} & OB & Yes & Fixed & $8.3<R/\text{kpc}<14$ $|z|<1$~kpc\\
        \citet{Evolution_disc_shape_Amores_2017} & 2MASS star counts & No & Fixed &  $R<18$~kpc\\
        \citet{Warp_RedClump_Lopez_Corredoira_2002} & Red Clump & Yes & Fixed & $R<13$~kpc\\
        \citet{Wang_LAMOST_Disk_Warp_RC_2020} & Red Clump & Yes & Fixed & $8.3<R/\text{kpc}<14$, $|z|<1$~kpc\\
        \citet{Chrobacova_GDR2_Warp_Model_2020GDR2} & Gaia DR2 & Yes$^*$/No & Fixed$^*$/Free & $R<20$~kpc\\
        \citet{Cheng_2020_Warp_patern} & K type stars & Yes & Fixed &  $R<16$~kpc\\
  \hline
	\end{tabular}
\end{table*}

We begin by comparing our results against those from \citet{Skowron2020}, obtained for the same Cepheid sample as used here. Within the uncertainties the two coincide at almost all radii. The \citet{Skowron2020} model behaves like an average smooth model around our results. The mean difference between both models for the northern region (for $R\gtrapprox10$~kpc) is $0.054$~kpc, and for the southern region is $0.043$~kpc. This level of agreement is expected because we are using a subset of their sample, the differences being in how we model the warp. \citet{Skowron2020} model the warp as a Fourier sum with $M=2$ (as we do) but assume a constant phase for each mode as a function of radius ($\partial_R\varphi_i=0$) and a second-degree polynomial for each amplitude ($A_m(R)=\gamma_m(R-R_d)^2$  where $\gamma_m$ is a constant) as a function of $R$. Under these assumptions the resulting model has the form 

\begin{equation}
    Z(\phi,R)=A_0+(R-R_d)^2\sum_{m=1}^2\gamma_m\sin(m\phi-\varphi_m).
\end{equation}

In consequence, there is a single Fourier sum that expresses the mean azimuthal behaviour of the warp at all $R>R_d$ which is scaled by the function $(R-R_d)^2$. In our model, without these assumptions, we can represent how the azimuthal geometry of the warp changes with the radius, giving rise to the differences between both models. The \citet{Skowron2020} model has the ability to reproduce the mean asymmetries observed in the warp, but not the LON twist or azimuthal changes in the different modes, which affect where the maxima are located.

We also compare our results with those from other warp models obtained for dynamically cold tracers like HI \citep{Levine2006}, Dust \citep{Marshall_Modelling_interstellar_extinction_2006}, OB stars \citep{Li_2023_Nonsteady_Warp}, Cepheid \citep{Chen3Dmap} and pulsars \citep{Warp_Pulsars_Yusifov_2004}. Because Cepheids are a young population \citep[$<500$~Myr, e.g.][]{CatelanSmith2015}, they are expected to still retain the warp shape inherited from the gas and its star-forming regions, so the agreement among young tracers is expected. We also show the results from \citet{Evolution_disc_shape_Amores_2017} selected for a young population with an age of $400$~Myr compatible with that of Cepheids. In the northern region, within uncertainties, we found excellent agreement with all previous results for young tracers, and a clear disagreement with results from \citet{Evolution_disc_shape_Amores_2017} inferred from star counts modelling using the Besan\c{c}on Galactic model. The warp model from pulsars departs the most from ours, with a mean difference of $0.14$~kpc (less than the intrinsic dispersion of Cepheids, see Fig.~\ref{fig:A_R}). For the HI model we found differences for $R<12$~kpc, which may be due to the underestimation by the amplitude fitted to its own results by \citet{Levine2006} between $10<R<12$. Compared to our results in the southern region, these works tend to underestimate the amplitude of the warp for $R\gtrapprox13$~kpc. The warp traced by pulsars underestimates the height the most, compared to ours, with a maximum difference of $0.42$~kpc. These differences may arise due to the symmetry imposed in the models for this radial range. The models from \citet{Levine2006}, \citet{Chen3Dmap} and \citet{Li_2023_Nonsteady_Warp} are strictly symmetric in this radial range, in consequence, the asymmetry given by the $m=2$ mode between both regions can't be represented. The difference with the model from \citet{Marshall_Modelling_interstellar_extinction_2006} may be due to its radial coverage which does not extend beyond $R\sim13$. 

Although the degree of agreement in the southern region is not as good as in the north, its clear that all young tracers follow a similar warp \citep{Chen3Dmap,Skowron2020,Li_2023_Nonsteady_Warp}. The clear exception to this agreement is the result from \citet{Evolution_disc_shape_Amores_2017}. Although the disagreement with the \citeauthor{Evolution_disc_shape_Amores_2017} results in the south is not as strong as in the north, they still found a  warp amplitude that is systematically lower than ours as well as all other works for Cepheids and similarly young tracers like dust and HI. 

We now focus our attention on the intermediate population: Red Clump stars \citep{Warp_RedClump_Lopez_Corredoira_2002,Wang_LAMOST_Disk_Warp_RC_2020}, A type stars \citep{Judith_AstarsWarp_2023}, K type stars \citep{Cheng_2020_Warp_patern} and the full Gaia DR2 population \citep{Chrobacova_GDR2_Warp_Model_2020GDR2}. Results from \citet{Warp_RedClump_Lopez_Corredoira_2002} in the radial range $R\lessapprox13$~kpc spanned by its observations (thick part of the line) shows agreement with our results and, as with the young populations, the agreement is better for the northern region. However, extrapolating this warp model (thin part of the line) for the outer region of the disc would yield increasing differences that would grow up to the order of a few kpc. Also, the models by \citet{Cheng_2020_Warp_patern} and \citet{Wang_LAMOST_Disk_Warp_RC_2020} for a 1-3Gyr population are in agreement within uncertainties for the northern region in $R\lessapprox12$. In the southern region both models are in agreement with our results for $R\lessapprox11.5$~kpc, after this radius the differences increase up to several kpc in the outer regions. The warp model presented by \citet{Chrobacova_GDR2_Warp_Model_2020GDR2} is in clear disagreement in both the northern/southern regions with all other warp models using similarly old tracers (like \citealt{Cheng_2020_Warp_patern}) and with ours and all other results for young tracers. As we will discuss in Sec.~\ref{sec:Disc_TimeEvol}, these differences in amplitudes between the models will become important in the determination of the pattern speed of the warp. Results from \citet{Judith_AstarsWarp_2023} for the kinematics of A-type stars population have shown a clear signal of the warp in the anticentre direction ($\phi=180$), the increasing vertical velocity as a function of the radius from $R\approx12$~kpc, reaching $\approx 6-7$~\kms at $R=14$~kpc, similar to our results.\footnote{Because \citet{Judith_AstarsWarp_2023} do not present a model of the warp traced by the A-type stars we can not include it in Fig.~\ref{fig:NScomparison}.}

The issue of the warp's dependence on age of the stellar tracer remains an open question. Older stellar populations like RGB stars, Red Clump stars and other tracers older than Cepheids may trace a similar warp considering the uncertainty in the parameters of each model and their validity range. Also, \citet{Cantat_Gaudin_Anders_2020_Painting_a_portrait} reported that stellar clusters typically \emph{older} than $1$~Gyr trace the southern region of the warp similarly to the Cepheids. Thus, it is unclear whether there are significant discrepancies between the warps traced by older and younger populations.

 Among previous results available at present, either the predictions of models with age dependency deviate significantly from the warp observed for bona fide young tracers like the Cepheids, as seen in the case of \citet{Evolution_disc_shape_Amores_2017}, or there is not enough discrepancy in the differences (considering uncertainties) to determine an age dependency for the warp, as in the case of results from \citet{Wang_LAMOST_Disk_Warp_RC_2020} shown in Fig~\ref{fig:NScomparison}. The uncertainties in the parameters obtained by \citet{Wang_LAMOST_Disk_Warp_RC_2020} for all ages are large enough to allow for the agreement of all models from 1 to 12 Gyrs with our result with Cepheids. In particular the model for 9 Gyrs (not shown), an age completely incompatible with that of Classical Cepheids, is the one in best agreement with our results. Taking into account the restrictions present in some of the models regarding the asymmetry and radial dependence of the warp, each model’s validity range in distance and azimuth, and the current precision of the observed warp using different tracers, it remains unclear whether or not there is an age dependency in the warp.

\subsubsection{Asymmetries and deviations from the tilted rings model}

Our results, as well as several previous ones, showed that a tilted rings model ($Z=A(R)\sin(\phi-\varphi(R))$ or $M=1$) does not explain many of the features observed at different radii in position and in kinematics. For $Z$, the presence of a plateau at $10\text{~kpc}\lessapprox R\lessapprox11$~kpc and  $\phi\approx180^\circ$ shown in the second panel of Fig.\ref{fig:Z_phi_3panels}, where the warp in $Z$ is already present, cannot be explained without an $m=2$ mode. At that distance the tendency of the disc to warp towards the southern hemisphere is clear at $\phi\approx240^\circ$, still far enough in azimuth from the strong obscuration towards the bulge ($|\phi|<90$) to be an effect of the selection function.The bootstrapping realisations shown in Fig.~\ref{fig:Z_phi_3panels} and Fig.~\ref{fig:Vz_phi_3panels} that the plateau is well recovered, for this reason we consider unlikely to be an artefact of statistical noise. The northern extreme lies in the first quadrant and so its inference is more affected by selection function effects due mainly to obscuration, hence, it is less well constrained than the southern extreme. Nevertheless, the extremes of the warp (in $Z$) are found to be $\approx120^\circ$ apart, while in a tilted rings model this difference must be $180^\circ$ by construction. The observed shape resembles the "S-Lopsided" warp model presented by \citet{RG19}. A better azimuthal coverage in the first quadrant and behind the bulge (currently unavailable due to extinction) would provide better constraints for this model. Our result is robust, however, since better coverage can only make the  difference between the warp extremes even smaller if the northern extreme lies closer to the bulge.

In the kinematics, a \emph{static} warp (i.e. $\omega_m=0$ and $\dot{A}_m=0$) with a plateau would create a distinctive shape in $V_z$. If we consider a star rotating with angular velocity $\Omega$ following the shape presented in the second panel of Fig.~\ref{fig:Z_phi_3panels}, then, because the star rotates in the direction in which $\phi$ decreases, after passing the minimum in $\phi\approx300^\circ$ the star increases its vertical velocity until it reaches the plateau ($\phi\approx180^\circ$) where $V_z\approx0$, then, on its way to the maximum $Z$ close to $\phi\approx60^\circ$ the star gains $V_z$ until a certain point after which its $V_z$ decreases to zero when it reaches the maximum $Z$. This creates two maxima in $V_z$, one before the plateau and another one after it. A toy representation of a plateau would be $Z(\phi)=A_1\sin(\phi)+\frac{A_1}{2}\sin(2\phi)$; for a star rotating with angular velocity $\Omega$ in a static plateau, this will give $V_z(\phi)=\Omega A_1(\cos(\phi)+\cos(2\phi))$, which shows the geometry described before. This shape is  observed in the second panel in Fig.~\ref{fig:Vz_phi_3panels}. We also take the ratio between the amplitudes of the modes $m=1$ and $m=2$ in $Z$ and $V_z$ and found consistency with what is expected from the toy model ($\frac{A_1}{A_2}\approx2$ and $\frac{V_1}{V_2}\approx1$ around $R\approx 10.5$). This peculiar signal was also observed in proper motions by \citet{RG19} in the RGB population, who interpret it was a signal of the lopsidedness of the warp. As we see here it is actually a characteristic signal of the S-Lopsided model due to its plateau.  An indirect evidence of the plateau is also illustrated in Fig.~\ref{fig:LON} by the large dispersion of the LON for $R\lesssim11.5$~kpc
where the LON is ill-defined. For $R>11.5$~kpc the plateau disappears, and the dispersion in Fig.~\ref{fig:LON} is sharply reduced as the disc is significantly inclined and the LON becomes well-defined.
For $R>11.5$~kpc other features that differ from a tilted rings model are still present, like the azimuthal asymmetry between the two extremes. The angular difference between them grows but never reaches $180^\circ$, meaning that an $m=2$ mode is needed to describe the galactic warp. In consequence the tilted rings (i.e. $M=1$) model is unable to accurately describe the observed azimuthal location of the warp extremes at any radius. 

In Sec.~\ref{sec:Asymmetries} we presented our results of the asymmetry between the north and south extremes in the Cepheid's warp. Asymmetry between the height of the warp extremes, or lopsidedness, has also been reported by \citet{Chen3Dmap} and \citet{Skowron2020} for the Cepheids sample, by \citet{Levine2006} for the HI component and also by \citet{RG19} for the OB and RGB populations. All these works seem to agree in the existence of an asymmetrical distribution, with the HI as the best exponent of this feature. In our results, the northern extreme is larger by 
$\approx0.25$~kpc at $11.5<R/\mathrm{kpc}<13$ which declines to a mean difference  $\approx0.1$~kpc for $R>13.5$~kpc as shown in Fig.~\ref{fig:Asym}. For comparison the figure also shows the North/South asymmetry for the warp model obtained by \cite{Skowron2020}. This difference behaves like a mean trend of our result as a consequence of the assumption of constant phases for the modes and the polynomial radial dependence of the amplitudes. 
The observed asymmetry in the outer disc is similar to that found for the OB population at $R\approx14$~kpc by \citet{RG19}, but note that \cite{RG19} report an amplitude for the warp traced by OB of $0.3$~kpc, much lower than the $0.8$~kpc we observe for the Cepheids warp. For the RGB stars, which are older than the Cepheids, \cite{RG19} report a larger asymmetry (red line) but with the opposite sign. This would mean the RGB present a warp with similar amplitude to the Cepheids but larger at the southern extreme. As we show in the following discussion, this may be due to an azimuth dependency of the asymmetry measured.

\begin{figure}
	\includegraphics[width=8cm]{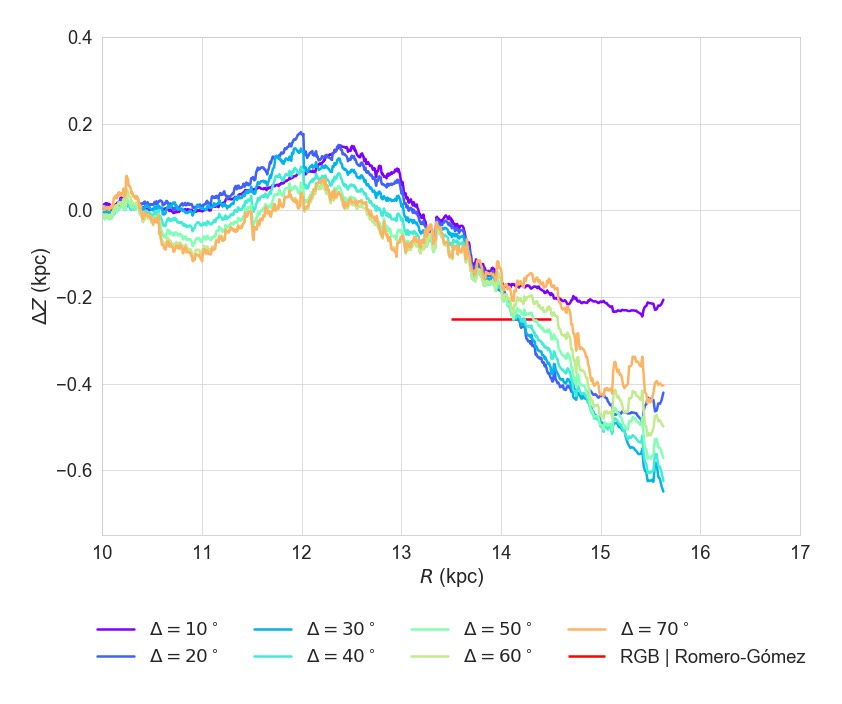}
    \caption{Difference $\Delta Z=|Z(180^\circ-\Delta\phi)| - |Z(180^\circ+\Delta\phi)|$ between two constant azimuths symmetric with respect to the anticentre direction $\phi=180^\circ$, as a function of galactocentric radius. Results for RGB stars reported by \citet{RG19} are also shown.}
    \label{fig:Delta_180}
\end{figure}

We have also found an azimuthal dependency in the asymmetry. Fig.~\ref{fig:NScomparison} shows how the warp at $\phi=90^\circ$ (north) and $\phi=270^\circ$ (south) presents a southern region with a larger departure from the midplane in the outskirts of the disc ($R\approx 15.5$~kpc), but in Fig.~\ref{fig:Asym} comparing the north and south extremes we get a northern warp height that is larger at all radii (see also Fig.~\ref{fig:XYlon}), with a decline in asymmetry towards a minimum almost constant value at the outer disc ($R>13$~kpc). Here, the twisted LON can create misleading interpretations in the measurement of the asymmetry, depending on how this measurement is made. Because the LON is leading and closely centred around $\phi\approx180^\circ$ (see Fig.~\ref{fig:XYlon}), a sample which covers the region $90^\circ<\phi<270^\circ$ will tend to cover mostly regions below the galactic plane (the mean $Z$ in azimuth between $90^\circ<\phi<270^\circ$ is below the plane for $R>13$~kpc). Hence, comparing $Z$ at symmetric azimuths  $\phi=180^\circ\pm\Delta$, rather than symmetric with respect to the LON, will tend to show a warp with larger amplitude below the plane. In Fig.~\ref{fig:Delta_180} we show the result of taking differences between the absolute values in height above/below the plane at lines of sight symmetric with respect to the Sun-Anticentre line, i.e. $|Z(180^\circ-\Delta)|$ and $|Z(180^\circ+\Delta)|$ for different $\Delta$. This clearly shows how measurements of the asymmetry at azimuths symmetric with respect to the Sun-Anticentre line  will yield a different result than when the extremes of the warp are compared. This is a consequence of both the twisted LON and the extremes of the warp never being diametrically opposed (Fig.~\ref{fig:Asym}). If the effect introduced by the twist in the LON found with Cepheids is also present in other populations, then different values for the asymmetry may not be enough to ensure different warps for different populations if the azimuthal dependency of the LON is not taken into account.

The results for the RGB sample obtained by \citet{RG19}, who reported a warp larger in the south than in the north, are also shown in Fig.~\ref{fig:Delta_180}. The north/south extremes found by \citet{RG19} are roughly symmetric with respect to the Sun-Anticentre line, so comparison in Fig.~\ref{fig:Delta_180} is appropriate, and this shows their results are consistent with ours for various $\Delta Z$ at their measured distance of $R\approx14$~kpc. Results for the OB population that were shown in Fig.~\ref{fig:Asym} to be in agreement with ours regarding the asymmetry are not shown here because they do not correspond to measurements made symmetric with respect to the Sun-Anticentre line.

\subsubsection{Line of nodes and line of maximum $V_z$}

In Sec.~\ref{sec:LON_LMV} we presented results for the LON. As was previously reported by \citet{Chen3Dmap} and more recently also by \citet{Dehnen_2023_Warp_Ceph}, the LON in the warp traced by Cepheids is twisted in the direction of the stellar rotation, meaning a leading LON, as shown in Figs.~\ref{fig:XYlon} and \ref{fig:LON}. This leading LON is in accordance with Briggs's Third Rule for warps \citep{Briggs_Rules_for_Galactic_WARPS}, which states that warp's LONs are straight for $R<R_{H_0}$ and twist for $R>R_{H_0}$, where $R_{H_0}$ is the Holmberg radius. Although these rules are derived for the warps traced by HI, they are expected to also apply for warps in the young population. \citet{Chen3Dmap} estimate the Holmberg radius at $R_{H_0}=11.4$~kpc, its LON and its $R_{H_0}$ are plotted in Fig.~\ref{fig:LON} (cyan curve and dashed vertical line, respectively). We found better agreement between the $R_{H_0}$ and our twist's starting radius, than that of \citet{Chen3Dmap}, which starts further out in the disc, as shown in Fig.~\ref{fig:LON}. For $R\gtrapprox 12.5$~kpc, the LON obtained by \cite{Chen3Dmap} is in quite good agreement with our results. We believe that the difference for $R<12.5$~kpc between both works is because, fitting only with one mode, the $m=1$ mode in \cite{Chen3Dmap} has to represent the whole warp despite its asymmetries. For this reason, and its relatively low amplitude in $R<12.5$~kpc, the  $m=1$ mode in \cite{Chen3Dmap} behaves as a mean between our LON (the full fit) and the phase of our $m=1$ mode (blue dots).  The LON twist is also suggested by \citet{RG19} to be present in the warp traced by RGB stars, but with an opposite direction, i.e. a trailing LON. However, \citet{RG19}  warn that this result may be driven by selection effects due to extinction.  

In Sec.~\ref{sec:LON_LMV} we show how the LON and the \LMV{} have a similar twist but they do not overlap, having an almost constant phase offset of $25\fdg4$ between them. Both lines lie in the region of the disc best populated by our data (as seen in Fig.~\ref{fig:XYlon}) and  best recovered in our tests with simulations from Sec.~\ref{sec:AzimutRadial}. Therefore, we consider both lines to be robust and not affected by biases. The difference in phase between the two lines could be due to the presence of $m\geq2$ modes in the overall warp. \citet{RG19} also found an offset between the LON and the maximum vertical proper motion for the RGBs and attributed it to the lopsidedness of the warp. According to \citet[][see their Fig. 8 and Sec. 5.1]{RG19} the \LMV\ for the RGBs may lie at $\phi\approx160^\circ-170^\circ$ (they observe $\mu_{b,LSR}$ rather than $V_z$), leading their LON (at $\phi\approx 180^\circ-200^\circ$) and also ours, but with a twist opposite to our results with Cepheids. Again, this result for the RGBs may be subject to selection effects which may have affected the inference of the LON.

In a warp dominated by the $m=1$ warp in both $Z$ and $V_z$, a change in amplitude with time could be responsible for the out-of-phase LON and \LMV. The phase offset $\delta$ between the LON and \LMV\ is given by $\varphi_1-\varphi_1^V+\pi/2$, so Eq.~\ref{Eq:Adot_n} translates into 

\begin{equation}
    \dot{A}_1=V_1\sin{\delta}  \label{eq:lonlmv_offset_Adot}
\end{equation}

\noindent directly relating the phase offset with the amplitude change. There are several caveats, however. First, as we have shown, for Galactic Cepheids the $m=1$ mode dominates the warp in $Z$ but not in $V_z$, in which the $m=2$ mode has a comparable amplitude at all radii. Second, an \LMV\ trailing the LON implies $\delta<0$, and Eq.~\ref{eq:lonlmv_offset_Adot} would require $\dot{A}_1<0$ in contradiction with our results and those from \citet{Dehnen_2023_Warp_Ceph}  shown in Fig.~\ref{fig:Adot}, which show that $\dot{A}_1\approx0$ up to $R\sim14$~kpc and $\dot{A}_1>0$ at larger radii. Therefore the evolution of the $m=1$ mode alone cannot explain the observed phase offset between the LON and \LMV.

We tested whether shifting the disc mid-plane can move the LON to coincide with the \LMV. To do so we would need to shift the stars by $-240$~pc, the mean vertical height of the stars along the \LMV. This would be too big a shift compared to the typical uncertainties of the position of the Sun above the Galactic plane \citep[of the order of tenths of pc, see e.g.][]{ZsolarChen}, making this explanation unlikely. 

In conclusion, we find the most plausible explanation for the phase offset between the LON and \LMV{} to be the presence of $m>1$ modes which deviate the \LMV{} from the LON, meaning that lopsidedness would \emph{indeed} be the main driver of this offset as suggested also by \citet{RG19} for the RGB sample. Samples with larger azimuthal coverage and also with measured line of sight velocities may help to confirm modes with higher frequencies in $Z$ and $V_z$, and settle the reason behind this out-of-phase LON and \LMV{}.

\subsubsection{The twist and velocity arcs}\label{sec:twist_and_arcs}

In Sec.~\ref{sec:LON_LMV} we showed the LON and \LMV\ are twisted and found these are well represented by straight lines in the plane $\phi,R$ (Eq.~\ref{Eq:3LONfit} for the LON and plus an offset for \LMV). These parameters are also explored by \citet{Dehnen_2023_Warp_Ceph} who present two LONs as a function of guiding radius for the warp traced by Cepheids, obtained from two different methods (mean orbital plane and mean position plane) and find a rate of change in the LON of $-14.7\pm 0.7\text{ deg}/{\text{kpc}}$ (mean orbital plane) and $-10.6\pm 0.8\text{ deg}/{\text{kpc}}$ (mean orbital position). Our result $-12.7\pm 0.3\text{ deg}/\text{kpc}$ lies between the two values. 

\begin{figure*}
	\includegraphics[width=17cm]{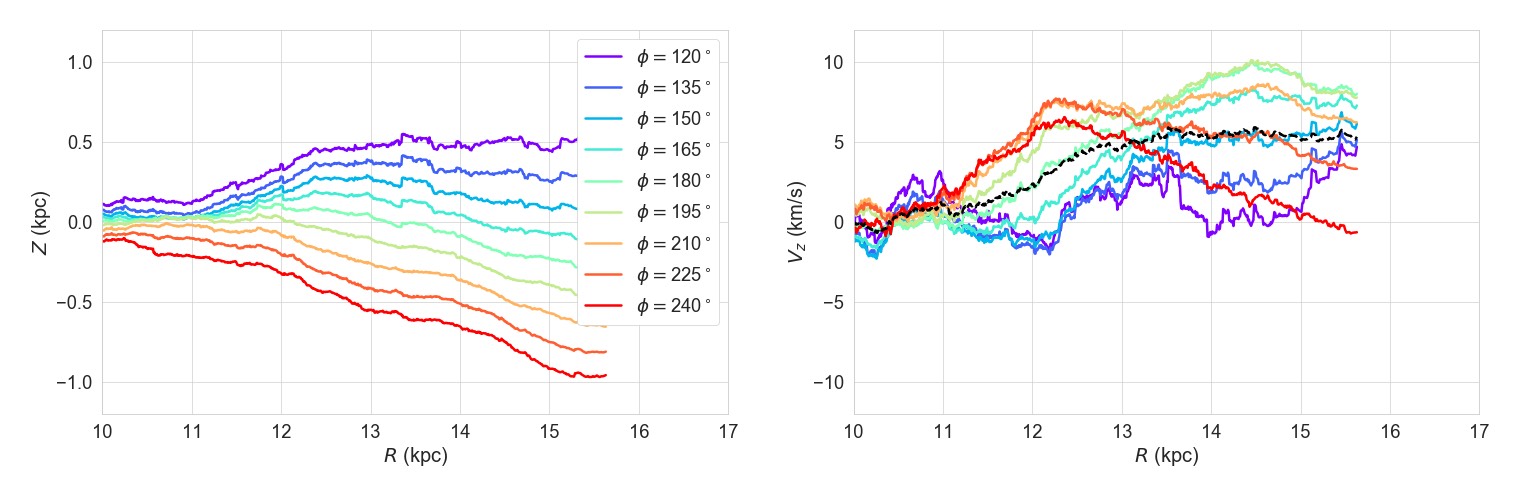}
    \caption{$Z$ (left panel) and $V_z$ (right panel) as a function of galactocentric radius for  nine evenly spaced azimuthal cuts from $\phi=120^\circ$ to $240^\circ$. The range of azimuth is selected to be in the region of the disc more populated by data and less affected by the SF. In the right panel the black curve represent the mean from the coloured ones} 
    \label{fig:V_zArc}
\end{figure*}

As the disc rotates and the LON wraps up and gets more and more twisted, the disc could appear to have ripples if the LON wraps around more than once around the disc.
The rate of change of the LON with $R$ can be associated with the inverse of a radial wavelength of these ripples. If we take a simple tilted rings model ($Z(R) \propto \sin(\phi-\varphi(R))$) and look at how it changes radially for a constant azimuth, e.g. $\phi=0$, then the warp will cross the plane at $\varphi(R_j)=j\pi$. Therefore, if the phase is described by $\varphi=\alpha R+\beta$, $\alpha$ it can be easily associated with a wavelength by $\lambda_\text{LON}=\frac{2\pi}{\alpha}$\footnote{In this approach we have ignored the amplitude ($\partial_RA/A\approx0$) which can change the distance between the peaks, but this change is negligible in comparison with our uncertainties}. This wavelength is the radial distance between two warp peaks at a constant $\phi$, if the LON and the amplitudes do not change its behaviour. For our LON we obtain $\lambda_\text{LON}\approx 28.4$~kpc. In the left panel of Fig.~\ref{fig:V_zArc} we show how this twisted LON creates long arcs in $Z$ for different azimuths. 

The right panel of Figure~\ref{fig:V_zArc} shows $V_z$ as a function of $R$ for different azimuthal cuts, in which the velocity is seen to create arcs whose peak changes in radius for different azimuths. These arcs are explained by the twisted \LMV\ together with the growth in amplitude in the kinematic signal as a function of $R$. These arcs are a direct consequence of the twist in the \LMV\ because the kinematic signal does not decline, and also because the peaks of the arcs move outward as phi decreases, as expected for the leading \LMV. Of course the change in amplitude and the asymmetries play a role in the position of the maximum, but the main driver of this arcs is the twist in the \LMV.  


These arcs in $V_z$ have been observed by previous works using Gaia DR2 and DR3  with other stellar tracers \citep{Cheng_2020_Warp_patern,GEDR3_anticentre_2021}. \citet{Cheng_2020_Warp_patern} pointed out that these arcs in $V_z$ are a consequence of the pattern speed in a tilted ring model, and indeed an arc can be created with just a constant pattern speed and a growing kinematic signal without a twisted LON. This is because the growing amplitude gets modulated by the factor ($\Omega-\omega$) so $V_z$ grows and then starts to decline as the co-rotation radius is achieved where $V_z$ is null (if $\dot{A}=0$), creating an arc. But this explanation can't take into account the change of the arc shape as a function of the azimuth (as shown in the right panel of Fig.~\ref{fig:V_zArc}), which can only be due to the twisted \LMV, which is a consequence of the twisted LON and the combination of the different evolutionary terms of the warp modes, not only the pattern speed of the $m=1$ mode.

Here we have shown \emph{these arcs in $V_z$ are a direct consequence of the twisted \LMV}. It is worth noticing that the \LMV\ does \emph{not track} the line of maximum in $V_z$ of the arcs presented in Fig.~\ref{fig:V_zArc} due to the change in amplitude as a function of $R$.



Because the \LMV{} has the same twist as the LON, $\lambda_{\text{LON}}$ represents also the radial distance between two $V_z$ peaks at a constant $\phi$. Using the same $\lambda_{\text{LON}}$ for the \LMV{} we may expect from a extrapolation of this oscillatory behaviour the minimum in $V_z$ at the anti centre direction to be around $R\approx28$~kpc and the point of null $V_z$ to be around $R\approx21$~kpc. \citet{Wang_2023_RotCurve30kpc} used Gaia DR3 to map the disc population out to $R\approx23$~kpc. In their Fig.~3 the disc's vertical velocity goes from positive to negative values at $R\gtrsim20$~kpc. These results seem to support our prediction, assuming $\lambda_{\text{LON}}$ holds for the entire Gaia DR3 sample used by \citet{Wang_2023_RotCurve30kpc}. Future extended maps of $V_z$ may prove helpful to explore whether this analysis also holds for $R>20$~kpc and for other stellar populations.  

Finally, \citet{Poggio_Measuring_vertical_response_2021} have shown with an N-body simulation of the Milky Way affected by the Sagittarius dwarf galaxy, that the $m=1$ mode has prograde rotation if the Milky Way disc and Sagittarius are close to an interaction. After approximately a few Myr the prograde motion coherent with the $m=2$ mode disappears. Perhaps the coherent rotation between the $m=1$ and $m=2$ modes close to an interaction found by \citet{Poggio_Measuring_vertical_response_2021} is the reason why the azimuth of the LON and the \LMV are well approximated by a monotonic (linear) dependence as a function of radius; without a coherent movement in the outskirts of the disc the LON may behave more erratically than is observed.

\subsection{Time evolution} 
\label{sec:Disc_TimeEvol}

In Sec.~\ref{Sec:TimeEvol} we provided a new formalism to derive the time change of each mode's amplitude ($\dot{A}_m$) and pattern speed ($\omega_m$) at each ring,  from the Fourier decomposition of its $Z$ and $V_z$. This new formalism is free from assumptions on how the amplitude and phase of each mode depends on the radius. By applying it to the Cepheids we derived the pattern speed and change in the amplitude of the $m=1$ mode for $R>12$~kpc.  

The dominant mode in $Z$ for the warp is the $m=1$ mode as expected for an S-type warp, so its evolution may drive most of the time evolution of the warp. In Fig.~\ref{fig:Omega_1} we show that, within uncertainties, $\omega_1$ shows a mean rotation of $9.2\pm 3.1$~km/s/kpc\footnote{The mean was obtained with measurements in independent rings for $R>12$~kpc} and its error corresponds to the standard deviation of posterior realisations from the independent rings, in agreement with previous reports from \citet{Poggio_Warp_evolving} and \citet{Cheng_2020_Warp_patern}, presented in Table.~\ref{tab:Omega_1}. Some oscillations are present, similar to the results obtained by \citet{Dehnen_2023_Warp_Ceph}, the main difference being that for $R<12$~kpc they found differential rotation slightly larger than we do, perhaps as a consequence of our use of the galactocentric radius as opposed to their use of the guiding centre. We note, however, our results from simulations (Appendix~\ref{app:validation_with_sims}) suggest $\omega_1$ may be overestimated in this radial range. 

\begin{table}
	\centering
	\caption{Pattern speed for the $m=1$ mode in (km/s)/kpc given by our mean value for $R>12$~kpc, \citet{Poggio_Warp_evolving} and \citet{Cheng_2020_Warp_patern}.}
	\label{tab:Omega_1}
	\begin{tabular}{lccr} 
		\hline
		 This work $\langle\omega_1 \rangle_{R>12\text{kpc}}$ & \cite{Poggio_Warp_evolving} & \cite{Cheng_2020_Warp_patern}\\
		\hline
		 $-9.18\pm3.12$ & $-10.86\pm3.23$ & $-13.57\pm0.2$\\
		\hline
	\end{tabular}
\end{table}

\citet{Chrobakova_Against_wapr_Precession_2021} present arguments about how the over-estimation in the amplitude of the warp leads to a over-estimation in its pattern speed\footnote{This relation holds for a warp with prograde rotation.}, therefore getting a lower amplitude of the warp will translate into a slower precession. This is well reflected by our Eq.\ref{Eq:omega_n}. However, the very low amplitude of the warp presented by \citet{Chrobacova_GDR2_Warp_Model_2020GDR2} seems unrealistic when compared to the rest of results from the literature, even compared to those with similar tracers as \citet{Cheng_2020_Warp_patern}. \citet{Chrobakova_Against_wapr_Precession_2021} also present a warp model for the younger population in its sample, with very similar results as obtained for the total sample. This particular disagreement in amplitude with the models of the young populations may indicate that the model by \citet{Chrobacova_GDR2_Warp_Model_2020GDR2} may be significantly underestimating precession rate of the warp, as a consequence of the underestimated amplitude. 

Equation~\ref{Eq:omega_n} also makes it clear why our results for $\omega_1$ are similar to those of \citet{Poggio_Warp_evolving} and  \citet{Cheng_2020_Warp_patern}, despite different assumptions in the three warp models, like the amplitude or the fixed phase $\phi_1$.
This equation shows that $\omega_1$ depends on the \emph{difference} between the phase of the mode in $Z$ and in $V_z$, therefore, it doesn't matter where they are located or if they are twisted, as long as the phase difference is the same. Because in the Milky Way the assumption that $\phi_1-\phi_1^V\approx-\frac{\pi}{2}$ seems to hold at least up to $R\approx14$~kpc, which is the same as assuming $\dot{A}_1\approx0$, independently of which $\phi_1$ the model adopts or if it is twisted or fixed $\omega_1$ will not be influenced by this assumption as long the model adopts $\dot{A}_1\approx0$. Also, the assumed amplitude affects $\omega_1$ as was previously mentioned, but $\omega_1$ gets saturated by over-estimations in $A_1$, because as $A_1\to\infty$ then $\omega_1\to\Omega$ (because the kinematic signature $V_1$ makes $\omega_1<\Omega$). This may be the reason why \citet{Cheng_2020_Warp_patern}, with its larger amplitude, gets a larger $\omega_1$ than ours, and also why \citet{Poggio_Warp_evolving} with the same kinematic signature gets a larger $\omega_1$, as it uses larger amplitudes. In this analysis of $\omega_1$ we have left $V_1$ constant because the kinematic amplitude of the warp seems similar for different tracers as shown by \citet{GEDR3_anticentre_2021}.
These could be the reasons why \citet{Poggio_Warp_evolving} and \citet{Cheng_2020_Warp_patern} get similar results to ours, even when they do not consider a twisted $\phi_1$ and when their amplitudes are larger than ours. We should add that the difference in amplitude is not the only parameter that plays a role in this analysis, the rotation curve and the kinematic signal are not the same between the works cited and they can change the pattern speed measurements, so we expect that these differences to also play a role.

Previous works on the time evolution of the warp neglect the contribution by the change in amplitude to the warp's kinematics \citep{Poggio_Warp_evolving,Cheng_2020_Warp_patern}. \citet{Poggio_Warp_evolving} argue that the effect of $\dot{A}_1$ may be a second order effect in the kinematics. Our results shows empirically that the change in amplitude can be neglected at least up to $R\approx15$~kpc. \citet{Wang_LAMOST_Disk_Warp_RC_2020}, finds the change in amplitude derived from the young population ($\approx1$~Gy) to be null, which within uncertainties is consistent with our mean measurement up to the radial limit to which \citet{Wang_LAMOST_Disk_Warp_RC_2020} restricted its sample, i.e. $R=14$~kpc. For $R>14$~kpc we found $\dot{A}_1>0$, reaching a maximum $\dot{A}_1\approx5$~\kms, this tendency is also observed in the change of the inclination in the tilted rings model by \citet{Dehnen_2023_Warp_Ceph} with similar values. 

The prograde rotation of the $m=1$ mode found with Cepheids is expected in the context of a disc embedded in a prolate halo as shown by \citet{Ideta_warps_in_prolate_haloes} and \citet{Jeon_2009_Warps_Triaxial_halos}. However, if this were the case, the prograde motion should be much slower ($0.1\text{ km/s/kpc}$ to $1.5\text{ km/s/kpc}$) than our result.  

Although the $m=1$ mode rotates almost rigidly, this does not guarantee a rigid rotation of the LON, because the $m=2$ mode also plays a role in the LON evolution, and in $V_z$ its amplitude is comparable to that of the $m=1$ mode. Due to the poor recovery expected for the $m=2$ mode (Sec.\ref{sec:RecOfModes}), a derivation of $\omega_2$ and $\dot{A}_2$ with our data would be biased, so we cannot ensure the evolution of the LON or of the whole warp to be one with rigid rotation. The $m=1$ mode also presents a growing amplitude for $R>15$~kpc, as is also reported by \citet{Dehnen_2023_Warp_Ceph}. For $R<14.5$~kpc the changes in amplitude are insignificant within the uncertainties, therefore, we present a warp which, at first order, shows a stable behaviour for $R<14.5$~kpc but still evolving in the outskirts of the disc.

In our derivation of $\omega_m$ and $\dot{A}_m$ we have ignored the radial velocity and azimuthal changes in $\Omega$. Considering a radial motion of $10$~km/s even when radial velocities may seem to be slower \citep{Cheng_2020_Warp_patern}, we found that $\dot{A}_1$ may change by about $1$~km/s and $\omega_1$ by $2$~km/s/kpc, which are in the order of the uncertainties. Also, that the radial bulk motion reported by \citet{Cheng_2020_Warp_patern} is inwards for $R\gtrapprox14$~kpc, will translate into a decrease in the measurement of $\dot{A}_1$ unless it is considered. Therefore, the growth in amplitude for $R\gtrapprox14$~kpc cannot be reduced by taking the radial motion into consideration (in fact, it should increase). These changes are smaller than the uncertainties in the results presented in this work, therefore we do not take them into account in our analysis. These features could be added to the analysis by considering a field of radial velocity and $\Omega$ described by Fourier sums at different radii. The extension of our formalism to account for the radial component will be presented in a future work.

\section{Conclusions}\label{sec:conclusions}

In this work we have used the \citet{Skowron2019} catalogue of Classical Cepheids to study the structure and kinematics of the Milky Way warp by means of Fourier Decomposition methods. These are the first results presented in the literature for the Fourier Decomposition of the warp in $V_z$. Our main results regarding the structure and kinematics of the warp are the following:

\begin{itemize}
    \item The warp is clearly lopsided, both in $Z$ and $V_z$. In $Z$, the amplitudes of the $m=1$ and $m=2$ modes are comparable up to $R\sim13$~kpc. At larger radii the $m=1$ mode dominates, as found previously by \citet{Chen3Dmap,Skowron2019}. In $V_z$, the amplitudes of the $m=1$ and $m=2$ modes are comparable at all radii. The $m=0$ mode does not play a major role in the overall warp shape, we detect  a bowl-like shape in the radial range $11.5<R/\text{kpc}<13$ with a maximum amplitude of $\approx 200$~pc. In $V_z$ the $m=0$ mode is almost null for $R>10$~kpc.  
    \item The warp presents a plateau at $10<R/\text{kpc}<11$. The observed shape resembles that of the S-lopsided model from \citet{RG19}. The double peak observed in $V_z$ at this radius is a kinematic signal associated with this plateau. It has also been observed in the proper motions of Red Clump stars by \citet{RG19}.  
    \item The warp is clearly asymmetric up to $R\sim13$, with a Northern warp larger than the Southern warp. In the outer disc ($R\gtrsim13.5$~kpc) the warp becomes symmetric to within uncertainties.
    \item The extremes of the Cepheid warp in $Z$ are never diametrically opposed. The difference in azimuth between the warp extremes is $\sim120^\circ$ at $R\sim10-11.5$~kpc and increases up to $140^\circ$ at $R\approx12.5$~kpc, remaining constant at larger radii. 
    \item The LON begins to twist at around $R\approx11$, which is close to the Holmberg radius for the Milky Way ($11.4$~kpc, \citealt{Chen3Dmap}), in agreement with Briggs' rules \citep{Briggs_Rules_for_Galactic_WARPS}. The LON's azimuth follows a linear relationship with radius, presented in Equation~\ref{Eq:3LONfit}.
    We found a twist of $-12.7\pm0.3\frac{\circ}{\text{kpc}}$.
    \item The \LMV{} does not coincide with the LON, but trails behind it with a constant offset of $25\fdg4$. We rule out that this offset is due to the change in amplitude with time of the $m=1$ mode, and explain this offset as a consequence of the lopsidedness also present in the kinematics.
    \item The arcs in $V_z$ as function of $R$ observed in other stellar populations \citep{Cheng_2020_Warp_patern,GEDR3_anticentre_2021} are also present in the Cepheids sample. We show these are a consequence of the twisted \LMV{} (see Figure~\ref{fig:V_zArc}).
\end{itemize}

We have also introduced a new formalism (Section~\ref{Sec:TimeEvol}), based on the joint analysis of the Fourier series in $Z$ and $V_z$, from which the pattern speed and instantaneous change in amplitude for each individual Fourier mode can be derived.
By applying this formalism to the Fourier Decomposition obtained for the Cepheids in $Z$ and $V_z$, we derive the pattern speed and amplitude change of the $m=1$ mode as a function of radius. Our main results are as follows:

\begin{itemize}
    \item The $m=1$ mode shows a prograde differential rotation for $11<R\mathrm{(kpc)}<13$ with $\omega_1$ going from $\sim -20$\kmskpc\ at $R\sim10-11$~kpc to $-9.18$~\kmskpc\ at $R\sim13$. Our results from simulations, however, suggest $\omega_1$ may be overestimated in $11<R\mathrm{(kpc)}<13$ this radial range.
    \item The amplitude of the $m=1$ mode remains approximately constant, with $\dot{A}_1 \approx 0$ \kms for $R<14.5$~kpc. The amplitude change has a growing tendency for $R>15$~kpc, reaching $\dot{A}_1\approx 5$~\kms at $R\approx15.5$.
\end{itemize}

Thanks to the precise measurements from Gaia DR3 and distances from \citet{Skowron2020} to its sample of Cepheids, we can explore the complex signal of the warp in both its structure and kinematics. Future Cepheid samples with increased coverage in the first and fourth quadrants will contribute to better restrict the parameters of the warp. A better understanding of the warp kinematics is necessary to make more robust comparisons with simulations and with analytical models of its dynamics, which can lead to better constraints on the possible history of the warp and its role in the evolution of the Milky Way disc's dynamics. {Furthermore, the complexity revealed may not be unique to the Galactic warp, understanding it will help also understand warps in external galaxies.}

\section*{Acknowledgements}
The authors thank the referee for the clear and succinct review which has enriched the presentation of our work. MC and CM kindly thank Marcin Semczuk and Walter Dehnen for an early review of this manuscript and for sharing their results prior to publication. 
This research has been supported by funding from project FCE\_1\_2021\_1\_167524 of the Fondo Clemente Estable, Agencia Nacional de Innovaci\'on e Investigaci\'on (ANII)  and  project C120-347 of the Comisi\'on Sectorial de Investigaci\'on Cient\'ifica (CSIC) at the Universidad de la Rep\'ublica, Uruguay.
MRG acknowledge funding by the Spanish MICIN/AEI/10.13039/501100011033 and by "ERDF A way of making Europe" by the “European Union” through grant PID2021-122842OB-C21.
TA acknowledges the grant RYC2018-025968-I funded by MCIN/AEI/10.13039/501100011033 and by ``ESF Investing in your future''. This work was (partially) supported by the Spanish MICIN/AEI/10.13039/501100011033 and by "ERDF A way of making Europe" by the “European Union” and the European Union «Next Generation EU»/PRTR,  through grants PID2021-125451NA-I00 and CNS2022-135232. 
MRG and TA acknowledge the Institute of Cosmos Sciences University of Barcelona (ICCUB, Unidad de Excelencia ’Mar\'{\i}a de Maeztu’) through grant CEX2019-000918-M.
This work has made use of data from the European Space Agency (ESA) mission {\it Gaia} (\url{https://www.cosmos.esa.int/gaia}), processed by the {\it Gaia} Data Processing and Analysis Consortium (DPAC, \url{https://www.cosmos.esa.int/web/gaia/dpac/consortium}). Funding for the DPAC has been provided by national institutions, in particular the institutions participating in the {\it Gaia} Multilateral Agreement.

\emph{Software:}
    Astropy \citep{astropy2018},
    Matplotlib \citep{mpl},
    Numpy \citep{numpy},
    Jupyter \citep{jupyter2016}, and 
    TOPCAT \citep{Topcat2005,Stilts2006}
\section*{Data Availability}

The catalogue of Classical Cepheids used in this research is publicly available in \citet{Skowron2020}. The results from our best fitting models as well as posterior samples are provided in Tables~\ref{tab:results} and \ref{tab:result_posteriors}.



\bibliographystyle{mnras}




\appendix

\section{Mathematical definitions for the inference}
\label{app:Post_Sol}









In this section we expand on the definitions of various mathematical objects used in Sec.~\ref{sec:FDM}. We begin defining the vector $\vec{\mathbb{CS}}(\phi)$ as

\begin{equation}
    \vec{\mathbb{CS}}(\phi)=[1,\cos(\phi),\cos(2\phi),...,\cos(M\phi),\sin(\phi),...,\sin(M\phi)],
\end{equation}

we can write the matrix $\mathbf{A}$ as 

\begin{equation}
    \mathbf{A}=\sum_{i=1}^N\frac{\vec{\mathbb{CS}}(\phi_i)\otimes\vec{\mathbb{CS}}(\phi_i) }{\sigma^2_i},
    \label{Eq:App.matrixA}
\end{equation}

where $\otimes$ denotes the outer product, $\phi_i$ the azimuth of the $i$-th star and the $\sigma_i$ its dispersion defined as 

\begin{equation}
    \sigma_i^2=\sigma_{z_i}^2+\sigma_{ID}^2,
\end{equation}
where $\sigma_{z_i}$ is the uncertainty in $z$ and $\sigma_{ID}$ the intrinsic dispersion.


The vector $\mathbf{p}$ is defended as
\begin{equation}
    \mathbf{p}=\sum_{i=1}^N\frac{z_i}{\sigma_i^2}\vec{\mathbb{CS}}(\phi_i).
    \label{Eq:App.VecP}
\end{equation}






\section{Validation with Simulations}\label{app:validation_with_sims}

In this section we use a warped galactic disc simulation to analyse the performance of the method described in Sec.~\ref{sec:FDM} when applied to mock data. We analyse how observational errors and the selection function of the data affect the recovery of each mode's parameters in $Z$ and in $V_z$, the full Fourier sum, and the intrinsic dispersion in different regions of the disc.

\subsection{Structure and Kinematics}

For our warped galactic disc model (without observational errors or selection function) we use the test particle simulation of the Sine Lopsided warp from \cite{RG19}. This is an S-shaped warp modified from a simple tilted rings model to allow for an warp with an arbitrary asymmetry \citep[a 3D representation is shown in Fig. B1 in][]{RG19}. The warp is such that the height of the mean plane of the disc is given by

\begin{equation}
    \langle z(R,\phi)\rangle =  R\sin(\phi)\sin(\psi(R,\phi))
    \label{eq:sine_lopsided_z}
\end{equation}

where
\begin{equation}
    \psi(R,\phi;R_1,R_2,\alpha,\psi_{up},\psi_{down})=[A+B\sin(\phi)]f(R;R_1,R_2,\alpha)
\end{equation}

with $A=\frac{1}{2}(\psi_{up}+\psi_{down})$, $B=\frac{1}{2}(\psi_{up}-\psi_{down})$ and $f$ having the following expression 

\begin{equation}
     f(R;R_1,R_2,\alpha)= \begin{cases}
      0 \ \ \ \ \ \ \ \ \ \ \ \ \ \ \ \ \ \ \  R\leq R_1\\
      \big(\frac{R-R_1}{R_2-R_1}\big)^\alpha \ \ \ R_1<R<R_2\\
      1 \ \ \ \ \ \ \ \ \ \ \ \ \ \ \ \ \ \ \ R\geq R_2
     \end{cases}
\end{equation}

where \citet{RG19} set $R_1=10.1\ \text{kpc}$, $R_2=14\ \text{kpc}$, $\alpha=1.1$, $\psi_{up}=7.5^\circ$ and  $\psi_{down}=4.25^\circ$. These parameters were chosen so that they would represent a plausible model of the asymmetry observed in the Galactic warp.

For the test particle simulation the strategy followed by \citet{RG19} was to initialise test particles in a a flat disc relaxed in an \citet{Allen_Santillan_potential} Galactic potential, then warp the potential adiabatically for five periods of the circular orbit at a radius $R=R_2$ and finally 
let the stars relax for a further two periods (at $R=R_2$).
The resulting configuration is such that stars at $R<R_2$~kpc are in statistical equilibrium with the imposed potential and their mean $z$ is described by Eq.~\ref{eq:sine_lopsided_z}.

The test particles were initialised with a vertical velocity dispersion of $16.6$ \kms representative of Red Clump stars at the solar radius. In this work we will apply the method to a sample of Cepheids, a kinematically colder population in which the detection of the warp is more favourable. The results obtained for the Red Clump simulation will, nonetheless, still be useful to understand the general advantages and flaws of the method described in Sec.\ref{sec:FDM}. 

The mock catalogue from \cite{RG19} includes the simulation of the Gaia DR2 observational errors and selection function (hereafter SF), as described in their Appendix D. In what follows we will use this mock catalogue down-sampled to $N_{tot}=\Ntot$ to match the number of Cepheids in our final catalogue (Sec.~\ref{sec:Sample}), keeping the simulated errors in proper motion, and assuming a $3\%$ error in distance, representative of the photometric distances for the Cepheids in our sample (Sec.~\ref{sec:Sample}). Errors for $V_z$ were propagated from distance and proper motion errors, assuming the radial velocity is inferred from the rotation curve, as described in Sec.~\ref{sec:Sample}. 

Since the test particles in the \citet{RG19} simulation are only relaxed up to $R=R_2=14$~kpc, we need to establish a ground truth model representative of the whole disc that can be used as a fiducial model against which results for mock catalogues are compared. We take as our ground truth model (from now on \GT{} model) the Fourier fits for $Z$ and $V_z$ obtained by applying the method described in the previous section to an arbitrarily large sample of the simulation, without errors or SF and fitting up to the second order mode ($M=2$)\footnote{We tested a larger $M$ and find that $M>2$ does not improve significantly our results.}. Other combinations for $N,N_{tot},M$ for the \GT{} model were tested and we find that the selected one optimises the computational time required and gives us the detailed information needed. 
Finally, we will call the SF model the Fourier fits obtained by applying the method to the mock catalogue affected by observational errors and selection function. In all cases, to compute $\sigma_{ID}$ we divide each ring in $15$ equally spaced azimuthal bins and find the weighted average of the standard deviation.

\begin{figure*}
	\includegraphics[width=17cm] {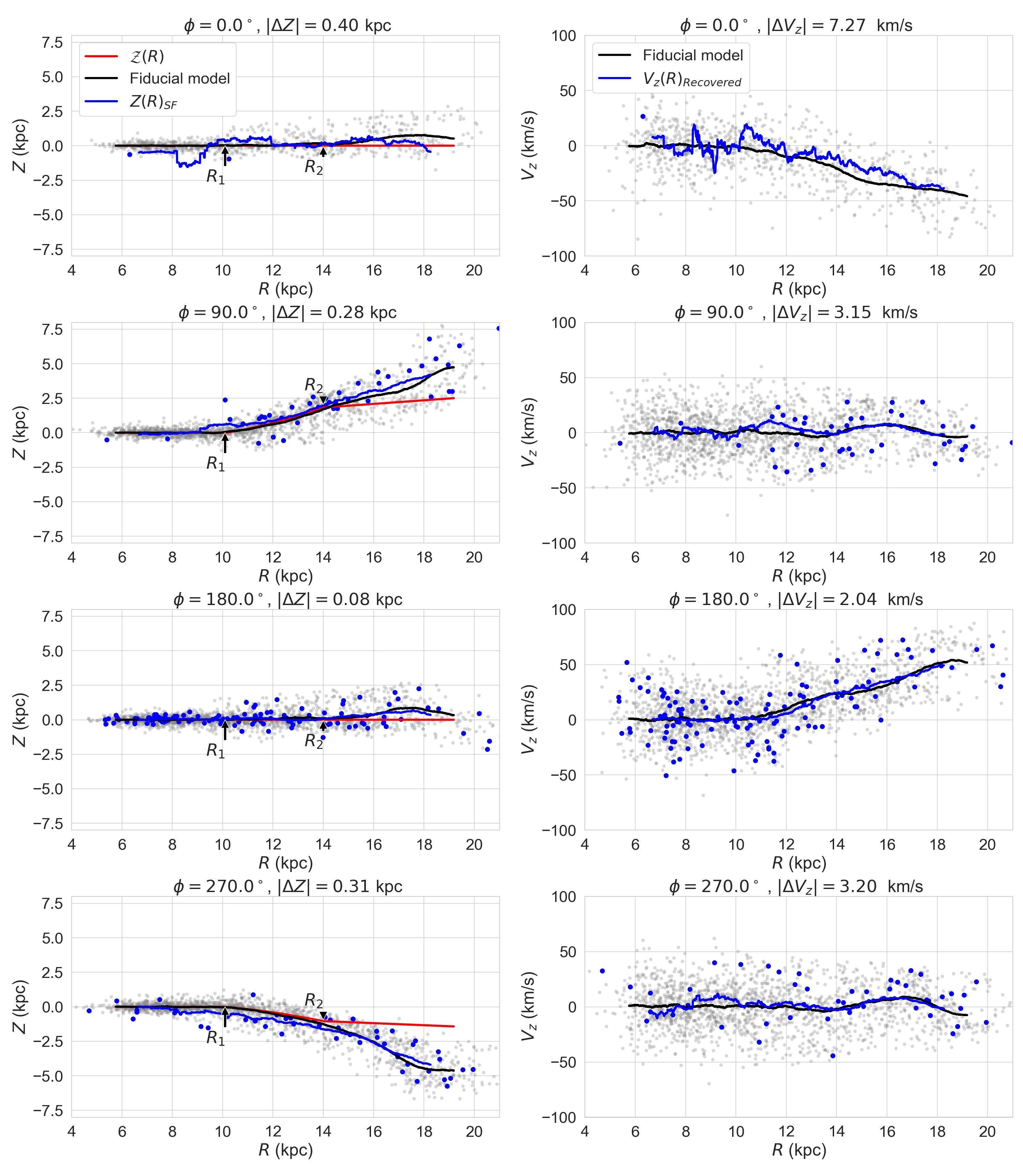}
    \caption{Each panel in the left column shows $Z$ as a function of $R$ centred at four different Galactocentric azimuths (from top to bottom $\phi=0.0^\circ$, $90.0^\circ$, $180.0^\circ$ and $270.0^\circ$) with $\Delta\phi=5^\circ$. The same is shown in the right column for $V_z$. The different solid curves show: the analytical prediction for the mean position of test particles in the warped potential (\Potmin{} curve), the ground truth model (\GTmodel{} curve) and the selection function model (\SFmodel{} curve) for the variable $Z$. The gray dots represent the stars used for the \GT{} model, and the blue ones represent the star used for the \SF{} model. The mean absolute difference between the \SF{} and \GT{} models for each $\phi$ is reported in the title.}
    \label{fig:GT_Z} 
\end{figure*}

Figure~\ref{fig:GT_Z} compares the distribution of test particles in the \GT{} (gray dots) and \SF{} (\SFmodel{} dots) simulated samples to the resulting \GT{} and \SF{} models (\GTmodel{} and \SFmodel{} lines) in $Z$ versus $R$ plots for three different azimuths. The figure shows in all cases the \GT{} model does indeed capture the behaviour of the test particles in the full radial range and coincides with the analytical prediction for $R<R_2$. Beyond this radius the stars cease to be in equilibrium with the potential and the mean $\langle z \rangle$ traced by the stars is not expected to follow Eq.~\ref{eq:sine_lopsided_z}. 
The \SF{} model agrees very well with the \GT{} model over the whole disc, capturing the overall behaviour of the warp. 
The mean absolute difference between the \SF{} and \GT{} models for each $\phi$ is reported in the top label of each panel in Fig.~\ref{fig:GT_Z} ($\langle|\Delta Z|\rangle_R$), the maximum $\langle|\Delta Z|\rangle_R\approx 0.40$~kpc corresponding to the region most affected by the bulge extinction ($\phi=0^\circ$), and the minimum $\langle|\Delta Z|\rangle_R\approx 0.08$~kpc which corresponds to the region towards the anticentre ($\phi=180^\circ$). In Fig.~\ref{fig:GT_Z} we notice that the general trend of the warp is recovered in all directions for the external region of the disc ($R\gtrapprox 10$) where the warp begins.

The right panel of Figure~\ref{fig:GT_Z} shows the corresponding results for $V_z$. In this case, we compare only the results for the \GT{}  and the \SF{} models, since there is no simple analytical form for $V_z(R)$ as discussed in Appendix~C in \citet{RG19}. Again, as in $Z$, the best-recovered region is around $\phi=180^\circ$ because it is less affected by the SF. All differences between the \GT{} and \SF{} models are much smaller than the corresponding velocity dispersion, which has a mean of $19.2$~km/s throughout the disc.
For both $Z$ and $V_z$ the reduced chi square $\chi^2_nu$ shows that the \GT{} model fits for $Z$ and $V_z$ are good ($\chi^2_\nu\approx 1$ $\forall\ R$). 

\subsubsection{Azimuthal and radial biases}
\label{sec:AzimutRadial}

\begin{figure*}
	\includegraphics[width=17cm]{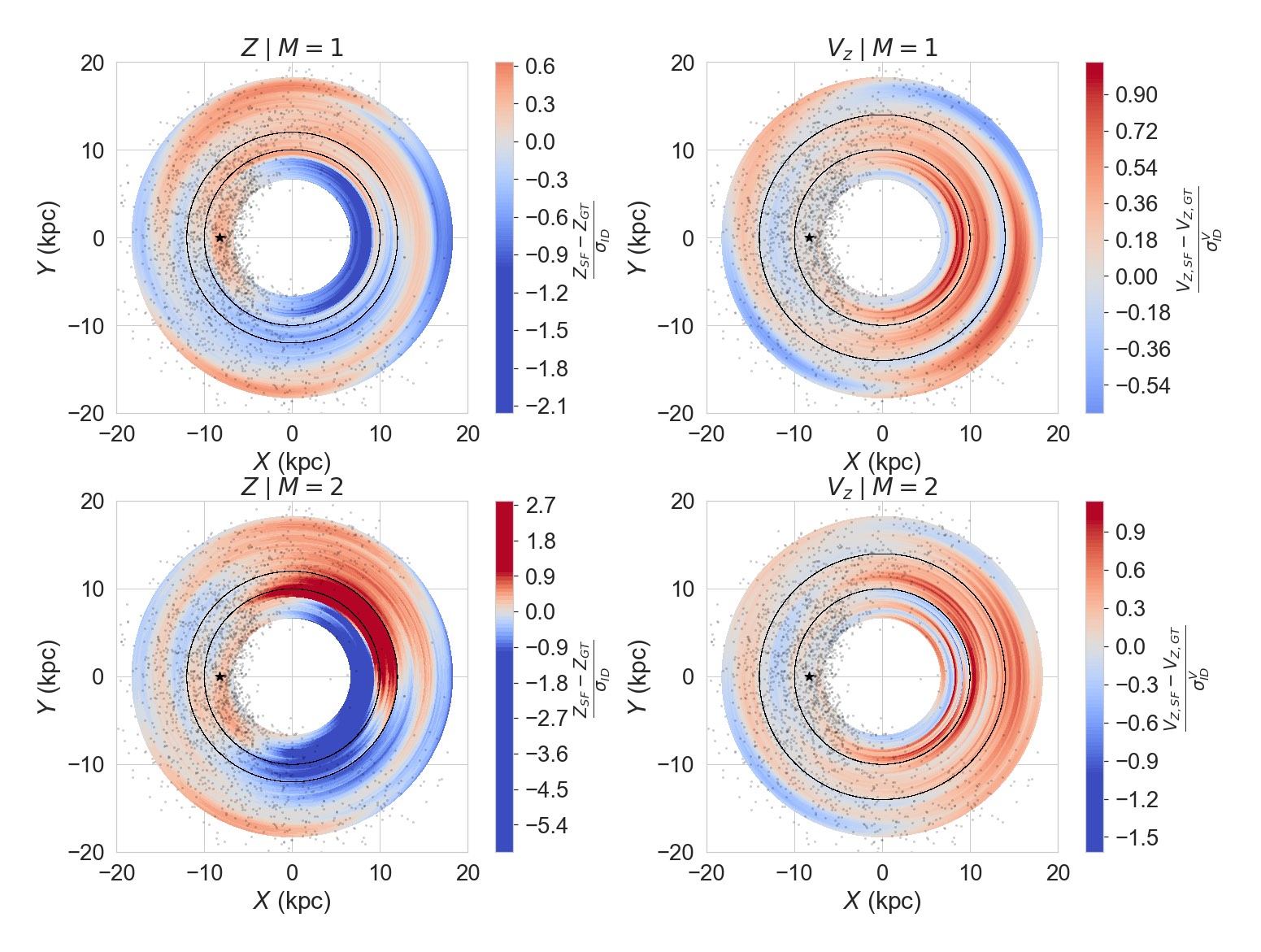}
    \caption{ Each plot shows the residuals between the fiducial model ($_{GT}$) and the model recovered with the mock catalogue ($_{SF}$), normalised by the intrinsic dispersion of the variable obtained in the fiducial model. The left and right panels show the residuals for $Z$ and $V_z$, respectively, for  $M=1$ (top) and $M=2$ (bottom). In each of the plots the inner black ring at $R=10$~kpc indicates where the warps starts, the outer ring is where the  $m=1$ mode begins to be greater in amplitude than the intrinsic dispersion. This happens at $R=12$~kpc for $Z$ and $R=14$~kpc for $V_z$. The pale gray dots are the stars in the mock catalogue used in the fit, and the black star indicates the  position of the Sun.}
    \label{fig:DeltaModel}
\end{figure*}

Figure \ref{fig:GT_Z} hinted the existence of regions in which the reconstruction of the warp given by the \SF{} model lacks accuracy. We argue this is due to the correlation between modes introduced by not having a uniformly distributed sample in azimuth and by stochastic clumps in regions with fewer stars in the sample due to the SF. To illustrate this, Figure~\ref{fig:DeltaModel} shows, in each panel, a residual plot between the \SF{} and the \GT{} model in $Z$ and $V_z$ (respectively top and bottom) normalised by the intrinsic dispersion given by the \GT{} model (with fixed $M=2$), for SF model fits with up to 1 (left) and 2 modes (right). Grey dots show the \SF{} sample, the black star shows the Sun's position, the inner ring is $R=10$~kpc where the warp begins and the outer ring is the radius in which the amplitude of $m=1$ mode is bigger than the intrinsic dispersion in the variable. The red and blue colours correspond to over/under estimations by the \SF{} model respectively.

In $Z$, for both $M=1$ and $M=2$ the differences are greater at $X>0$~kpc  in the inner region before the warp begins. The discrepancy is larger for $M=2$ because when fitting with a higher number of modes, in the areas most affected by the SF the higher frequency modes tend to drive the fit towards the few data points available introducing spurious oscillations where there is less data. For both $M=1$ and $M=2$ the recovery is best for outer radii, where the warp amplitude is larger than the intrinsic dispersion. For $M=1$, the differences start growing with radius due to the simulated warp’s asymmetry which is not well represented by the Fourier series with $m \leq 1$ modes, generating an $m= 2$ pattern in the differences. By contrast, the asymmetry is better captured by the series for $M=2$ for which the discrepancies in the outer region are smaller. However, a hint of the $m=2$ pattern in the differences still remains; this is due to a lower amplitude of the $m=2$ mode recovered by the \SF{} model (this is illustrated in left panel of Fig.~\ref{fig:A_Sine_ZVz}). 



For $V_z$ we analyse the bottom panels in Fig.\ref{fig:DeltaModel}, the left plot for $M=1$ and the right for $M=2$. The differences between the \SF{} and \GT{} models are always smaller for $V_z$ than the intrinsic dispersion in the whole disc, both for $M=1$ and $M=2$, in contrast with the recovery in $Z$ where differences exceed the intrinsic dispersion in the inner region. The best and worst recovery for $V_z$ are found in the same regions as for $Z$ because the azimuthal distribution is the same for both samples; with the best recovery at negative $X$, and the worst in the internal disc at positive $X$. Finally, the differences between the \SF{} and \GT{} models are much lower in $V_z$ than in $Z$. As also discussed in \citet{RG19}, this is expected because the SF creates exclusion zones in $Z$ due to high extinction near the Galactic plane, but doesn't in $V_z$ because the correlation between $z$ and $v_z$ is weak for a given star.

Given these results, we decide to use $M=2$ for Fourier fits for this work because it offers the least biased recovery for the region of the disc where the warp is most prominent (i.e. outer radii). Reliable results for the inner region of the disc are limited to $90\gtrsim\phi\gtrsim270$~kpc, the region least affected by the SF with best coverage, where biases in the recovery are lowest.

\subsubsection{Recovery of individual modes}
\label{sec:RecOfModes}

So far we have analysed the recovery of the shape and kinematics of the warp as a whole, given by the sum of the $M$ individual modes in the Fourier series.
Now we will analyse how well each mode is recovered. 

\begin{figure*}
	\includegraphics[width=17cm]{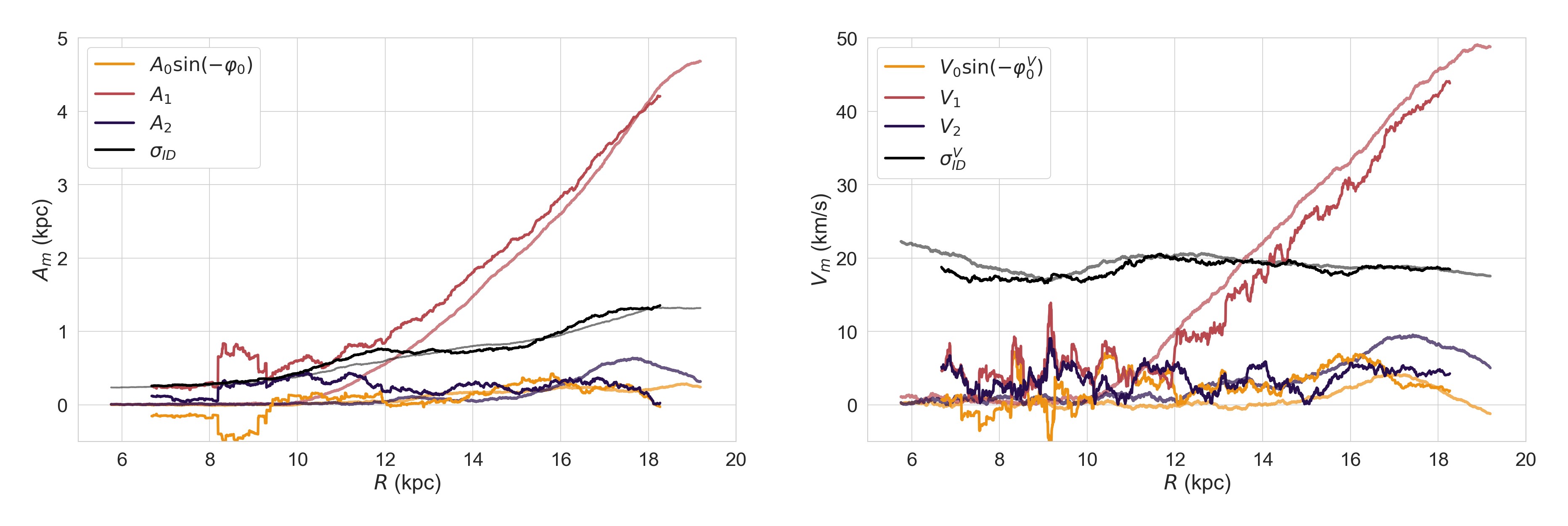}
    \caption{Each panel show the amplitude for the $m=0$ (yellow), $m=1$ (blue), $m=2$ (violet) modes and $\sigma_{ID}$ (black) for the \SF{} model (dark solid lines) and the \GT{} model (pale solid lines) obtained for $Z$ (left panel) and $V_z$ (right panel). Each amplitude and $\sigma_{ID}$ is plotted as a function of radii.}
    \label{fig:A_Sine_ZVz}
\end{figure*}

Each mode $m$ is characterised by its amplitude $A_m$ and its phase $\varphi_m$ in $Z$, and in $V_z$ with $V_m$ and $\varphi^V_m$. In Fig.~\ref{fig:A_Sine_ZVz} we compare the amplitudes for $Z$ (left) and $V_z$ (right) as a function of $R$ recovered for the \SF{} model (dark solid lines) against the values given by the \GT{} model (pale solid lines) for each mode. The intrinsic dispersion as a function of radius is also plotted in each panel.

The left panel of Fig.~\ref{fig:A_Sine_ZVz} shows how for inner radii ($R<10$~kpc) the disc is flat before the onset of the warp, as shown by the near zero amplitudes for all modes in the \GT{} model. Particularly for $m=1,2$, the \SF{} model finds non-zero amplitudes of the order of the intrinsic dispersion. Amplitudes are overestimated in the inner disc because the modes make the full Fourier series flat in the region less affected by the SF ($\phi\approx 180^\circ$), but it also tries to fit stochastic clumps far from the midplane at $\phi\approx 0^\circ$ where the SF has removed stars preferentially in the disc plane. At the outer parts of the disc, the $m=1$ mode is overestimated by the \SF{} model but the bias is reduced at the external part of the warp ($R>13$kpc) where the $A_1$ amplitudes are larger. The $m=2$ mode is overestimated due to correlations with other modes when the whole fit of the series is driven by stochastic clumps at $R\lesssim10$~kpc, as for the $m=1$ mode. The $m=0$ mode is well recovered over the whole disc.

Some features observed for the amplitudes in $Z$ are present also in $V_z$. For example the amplitudes are not $0$~\kms for $R<R_1$ due to the sparse azimuthal coverage caused by the SF. For $V_z$, the amplitude of $m=1$ is underestimated but the general trend is well recovered by the \SF{} model for $R>R_1=10$~kpc as in $Z$. The amplitudes of $m=0,2$ have differences between the \SF{} and the \GT{} model, also as in $Z$, which is expected because both amplitudes in the \GT{} model are smaller than $\sigma_{ID}$, which makes them harder for the \SF{} model to recover.  

\begin{figure*}
	\includegraphics[width=17cm]{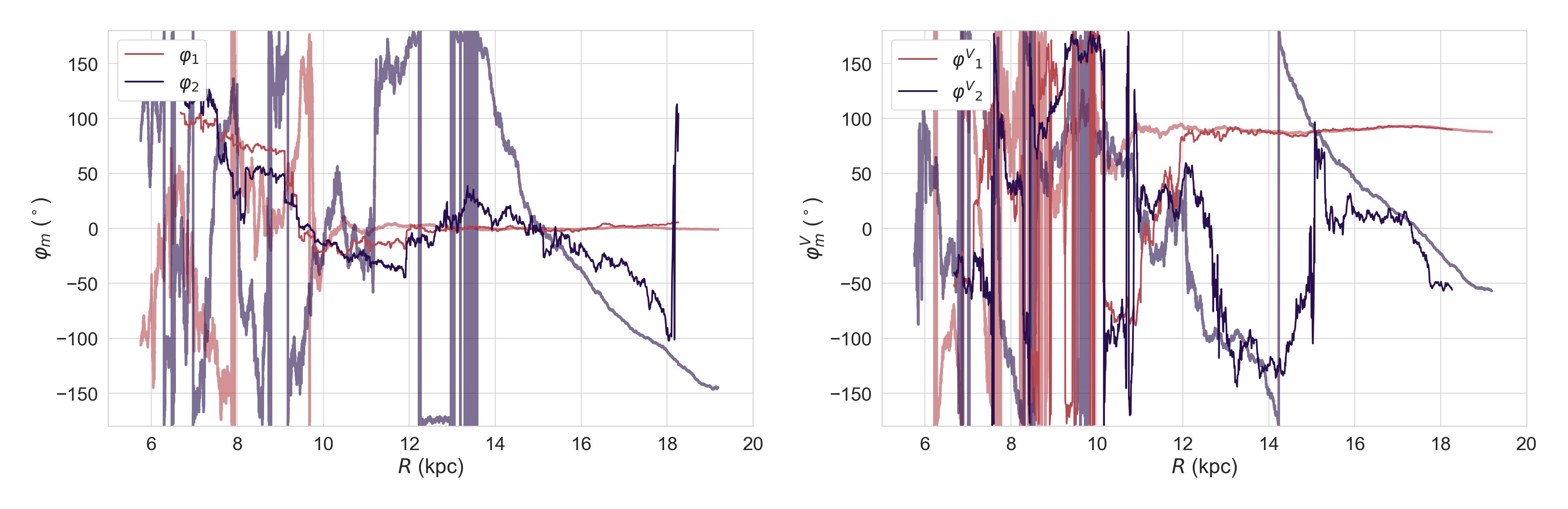}
    \caption{Each panel shows the phase for the modes $m=1$ (blue) and $m=2$ (violet) for the \SF{} model (solid dark line) and the \GT{} model (solid pale line) obtained for $Z$ (left panel) and $V_z$ (right panel), as a function of galactocentric radius $R$.}
    \label{fig:Varphi_Sine_ZVz}
\end{figure*}

Similarly to Figure~\ref{fig:A_Sine_ZVz}, in Figure~\ref{fig:Varphi_Sine_ZVz} we compare how the phase of each mode in $Z$ (left) and $V_z$ (right) is recovered by the \SF{} model as a function of radius. We do not plot the phase for $m=0$ because it can only take two possible values ($-90^\circ$ and $90^\circ$). 

For the inner disc at radii $R\lesssim R_1=10$~kpc before the onset of the warp, it is normal that the phase is badly recovered for all modes because the (true) amplitudes are near zero at these radii and the phase becomes meaningless. For the $m=1$ mode the phase for both $Z$ and $V_z$ are very well recovered, with no significant bias, for $R\gtrsim12$~kpc where $A_1>\sigma_{ID}$. For $m=2$, the general trends are recovered for $R\gtrsim12$~kpc, e.g. the twist in $Z$ and $V_z$ showing the change of phase as a function of radius.  However, we must be cautious in any particular analysis of $m=2$ as an individual mode due to the lack of recovery by the \SF{} model with this mode, its phase recovers some of its tendency but without accuracy.


\subsubsection{Intrinsic dispersion}

\begin{figure}
	\includegraphics[width=8cm]{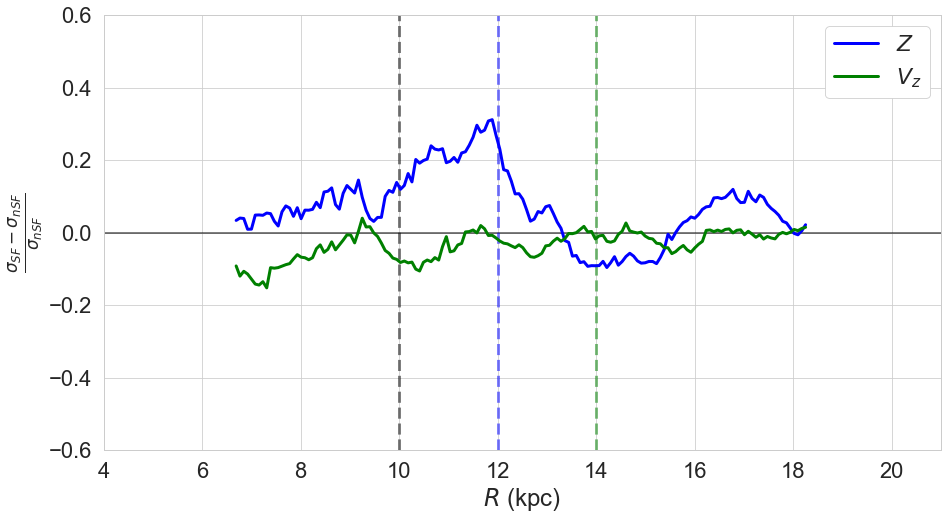}
    \caption{The ratio between the difference in intrinsic dispersion by the \GT{} model and the one obtained with the \SF{} model, over the intrinsic dispersion of the \GT{} model as a function of the galactocentric radii. The blue and green continuous curves correspond to the intrinsic dispersion of $Z$ and $V_z$, respectively. The black, blue, and green vertical dashed lines indicate the radius at which the warp begins, when the amplitude of the mode $m=1$ is bigger than the intrinsic dispersion for $Z$ and for $V_z$, respectively.}
    \label{fig:DeltaID}
\end{figure}

Finally, we analyse the bias introduced by the SF to the intrinsic dispersion that our method calculates. To do so, we compute the fractional difference between the $\sigma_{ID}$ obtained with the \GT{} and \SF{} samples. These differences for $Z$ (blue curve) and $V_z$ (green curve) are plotted in Fig.~\ref{fig:DeltaID} as a function of radius. The black vertical dotted line at $10$~kpc indicates the beginning of the warp, the blue one when the mode $m=1$ for $Z$ starts to be greater than $\sigma_{ID}$, the green one is the same as the blue but for $V_z$. 

First, for $Z$ the recovered $\sigma_{ID}$ is increasingly overestimated at inner radii until the warp becomes greater than the disc's thickness; for larger radii, the recovered $\sigma_{ID}$ decreases and is off just by $10\%$ of the \GT{} value. Both effects are due to the combination of the increased warp amplitude and the SF. The SF makes the stars near the plane very unlikely to be observed due to high extinction, while the stars away from the plane are less affected by it; since these stars are further away from the disc plane (because of the amplitude of the warp) this tends to inflate $\sigma_{ID}$ for $Z$. This effect is expected to be smaller for a dynamically colder stellar population like Cepheids. For $V_z$, on the other hand, we find a mean underestimation of $3\%$, much smaller than for $Z$. We find the appearance of the warp signal in $V_z$ has no effect in the ability to recover $\sigma_{ID}$. Overall the recovery of the intrinsic dispersion affects the inference on the amplitudes and phases in terms only of the dispersion of the posterior PDF, it does not introduce any systematic biases in the parameters themselves.

\subsubsection{Assumptions on the rotation curve}

We tested how the assumed rotation curve may affect the inference in the simulations and with the real data. We did not find any systematic bias in the amplitudes, phases and intrinsic dispersion inferred from the mock catalogues when we used the $v_z$ derived from the rotation curve, even when using different rotation curves. 

In the case of the Cepheids, we tested whether changing the rotation curve offset by $\pm10$~\kms could change our main results. We found that different offsets change the amplitudes of the $V_z$ arcs by $\sim1$\kms but do not change the general trend of the kinematic signal of the warp. The changes in $\omega_1$ and $\dot{A}_1$ due to changes in the rotation curve are insignificant in comparison with the uncertainties.




\subsection{Time Evolution}\label{sec:validation_time_evol}

In this Section, we validate the inference of $\omega_m$ and $\dot{A}_m$ by applying the formalism developed in Sec.~\ref{Sec:AdotOmega} to the simulated sample affected by the SF and comparing it to results for the \GT{} model (as in Sec.~\ref{sec:RecOfModes}). 
By doing this we're assuming that the formalism developed holds and will yield correct results for the \GT{} model. Since the test particle simulation we are using has a fixed warp, we expect from this test to recover a constant amplitude and null pattern speed in the region at equilibrium with the potential (i.e. $R<R_2=14$~kpc). In the outer parts the warp would be expected to evolve with time as the stars relax in the potential. Because the warp model used in the test particle simulation is not constructed by definition as a Fourier series it is not straightforward to use this data to test the recovery of specific values of the time evolution parameters. More involved tests in this direction could be done in a future work to validate the method.

In what follows we analyse the difference between the parameters from both models for $R>10$~kpc where the warp is present. We apply this formalism only to the $m=1$ mode due to the bias and noisy recovery in the $m=2$ mode parameters discussed in Sec.~\ref{sec:RecOfModes}.

\begin{figure}
	\includegraphics[width=9cm]{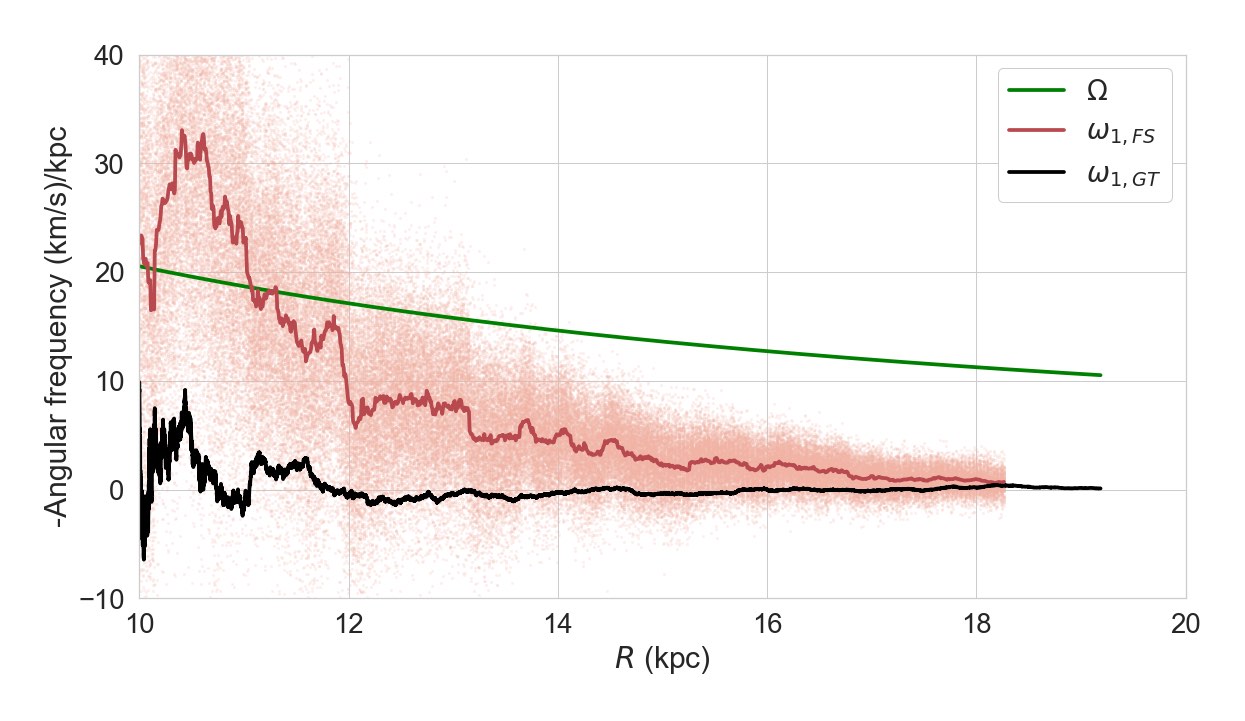}
    \caption{Angular frequency as a function of the galactocentric radii of the pattern speed $\omega_1$ for the $m=1$ mode, calculated from Eq.~\ref{Eq:omega_n} for the \GT{} model (black curve) and for the \SF{} model (red curve), the angular velocity of the stars $\Omega$ (green curve) is derived analytically from the \citet{Allen_Santillan_potential} potential. The red dots around each $\omega_1$ are $500$ realisation taken from the posterior at each ring for the \SF{} model.}
    \label{fig:omega_FS_GT}
\end{figure}

Figure~\ref{fig:omega_FS_GT} shows the results for the \GT{} and \SF{} models for $\omega_1$ as a function of radius. For the \GT{} model we get $\omega_1=0$ for $R>12$~kpc (black curve in Fig.~\ref{fig:omega_FS_GT}). 
The variations observed in $\omega_1$ at $10<R/\text{kpc}<12$ are expected in this region were the amplitude of the mode is still very low and it's pattern speed ill-defined. As the amplitude of $m=1$ mode increases the pattern speed recovered for the \SF{} model converges to results for the \GT{} at the outermost radii.  The mean overestimation in $\omega_1$ for $R>12$~kpc is of the order of $4$~\kms/kpc, which is within the uncertainties given by the posterior realisations (gray dots).

\begin{figure}
	\includegraphics[width=9cm]{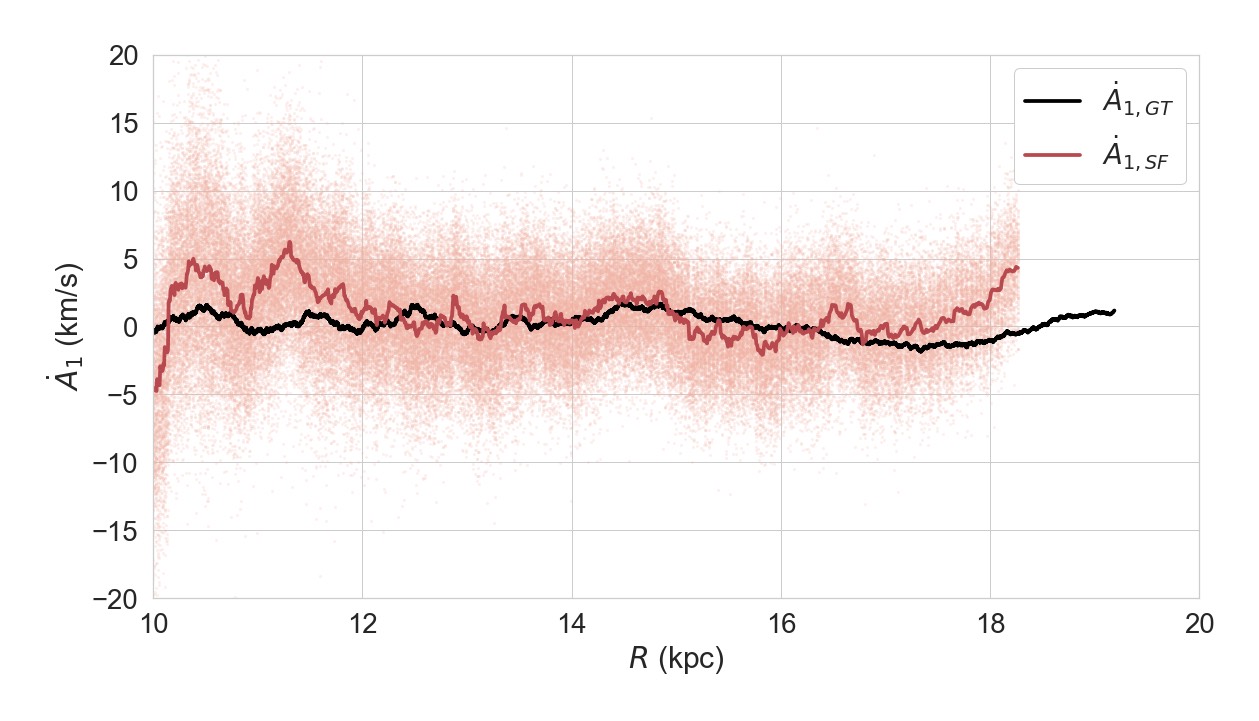}
    \caption{The change in amplitude $\dot{A}_1$ for the $m=1$ mode as a function of the galactocentric radius calculated from Eq.~\ref{Eq:Adot_n} for the \GT{} model (black curve) and for the \SF{} model (red curve). The red dots around each $\omega_1$ are $500$ realisation taken from the posterior at each ring for the \SF{} model.}
    \label{fig:Adot_FS_GT}
\end{figure}

Figure~\ref{fig:Adot_FS_GT} shows the result for $\dot{A}_1$ for the \GT{} (black curve) and the \SF{} models (blue curve). The difference between the two for $R<12$~kpc is due to the poor recovery in $\varphi_1^V$ as shown in Fig.~\ref{fig:Varphi_Sine_ZVz}. The mean difference for $R<12$~kpc between the recovery with the SF and GT models is less than $2$~\kms, which is within the uncertainties given by the posterior realisations (gray dots).
For $12<R/\text{kpc}<17$, the general tendency for $\dot{A}_1$ is recovered within the uncertainty with not appreciable bias. The relatively large uncertainty in the recovery on  $\dot{A}_1$ stems from small differences in  $\varphi_1-\varphi_1^V$, which near $\pi/2$ translate in large differences in $\dot{A}_1$ due to it dependence on the difference via a cosine function (Eq.~\ref{Eq:Adot_n}). The opposite happens for $\omega_m$ because it depends on the difference via a sine function.

\section{Goodness of fit and results for individual modes }
\label{app:ResultsIndividualModes} 

This appendix presents the results for the goodness of fits and summarises the results of Sec.~\ref{sec:Results} for the individual modes.

\subsection{Goodness of fit}
\begin{figure}
	\includegraphics[width=7cm]{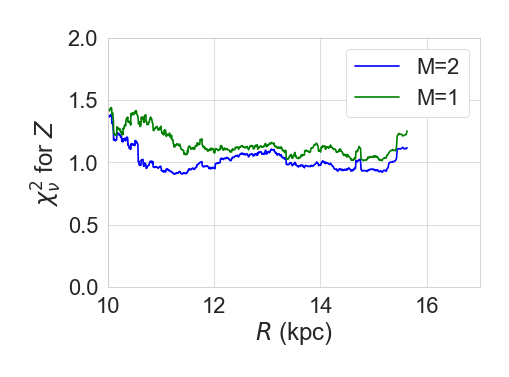}
    \caption{Reduced Chi-square for the fits in $Z$ done with $M=1$ (green curve) and $M=2$ (blue curve) as a function of galactocentric radius.}
    \label{fig:Chi_z}
\end{figure}

\begin{figure}
	\includegraphics[width=7cm]{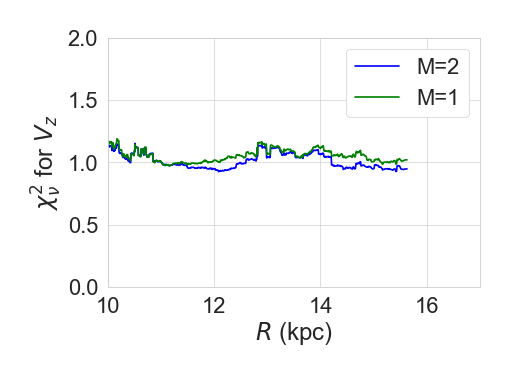}
    \caption{Reduced Chi-square for the fits in $V_z$ done with $M=1$ (green curve) and $M=2$ (blue curve) as a function of galactocentric radius.}
    \label{fig:Chi_Vz}
\end{figure}

We have tested with the reduced Chi-square how meany mode where needed to do the fits in $Z$ and $V_z$ for the Cepheids sample. Fig.~\ref{fig:Chi_z} and Fig.~\ref{fig:Chi_Vz} shows as a function of the radius the results for $Z$ and $V_z$. Clearly the result favoured the fits for $M=2$ for both variables, showing the need of the $m=2$ mode to reflect the asymmetries present in the warp. We have also computed the Bayesian information criteria (BIC, \citet{Ivezic_Statistics_2014}) for different radii and we found that $M=2$ is always clearly the best model for $Z$ at all radii $>10$~kpc. This is of special importance since $Z$ is more sensitive to biases due to the selection function problems. For $V_z$ the fits with $M=2$ is also the best case for the outer disc where the amplitude of the warp is significant. We have therefore chose the $M=2$ model for both variables.

\subsection{Individual modes}

\begin{figure*}
	\includegraphics[width=17cm]{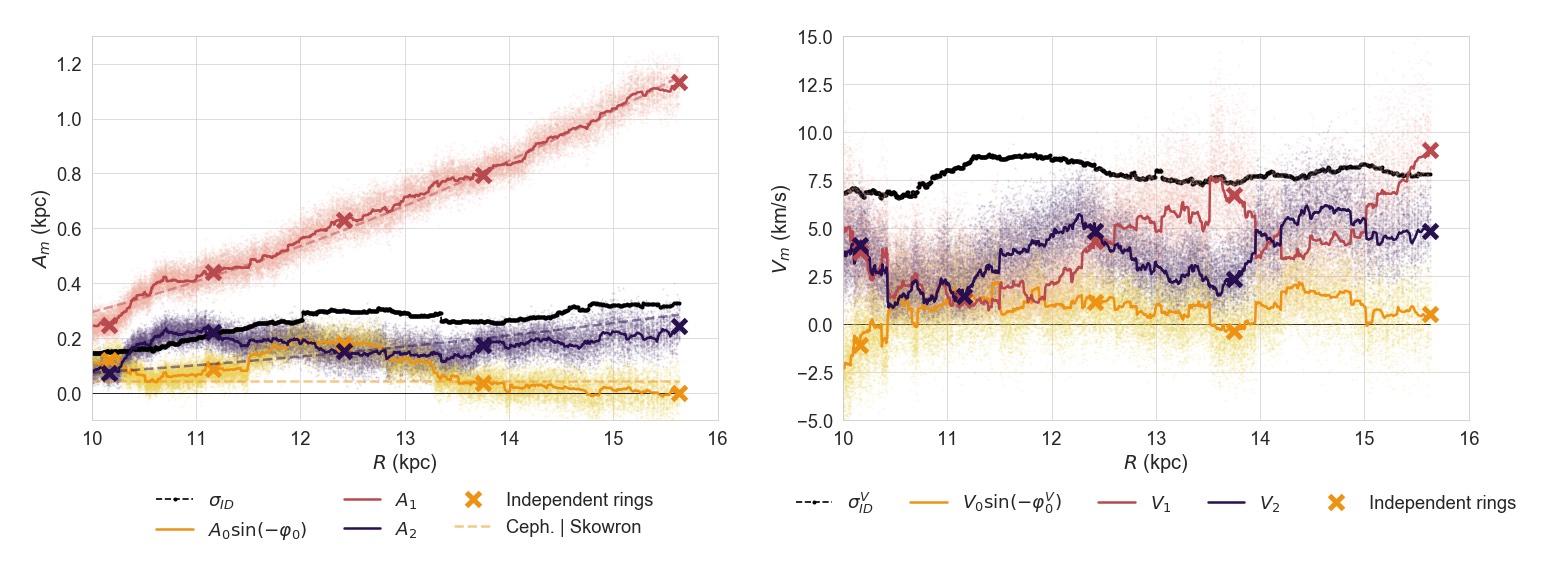}
    \caption{Amplitudes of each mode for $Z$ (\textbf{left} panel) and $V_z$ (\textbf{right} panel) as a function of galactocentric radius. The black curve shows the intrinsic dispersion for each radius in the respective variable. The dotted line for the amplitudes in $Z$ shows the results from \citet{Skowron2020}. The colour dots around each mode are $500$ realisation taken from the posterior at each ring.}
    \label{fig:A_R}
\end{figure*}

\begin{figure*}
	\includegraphics[width=17cm]{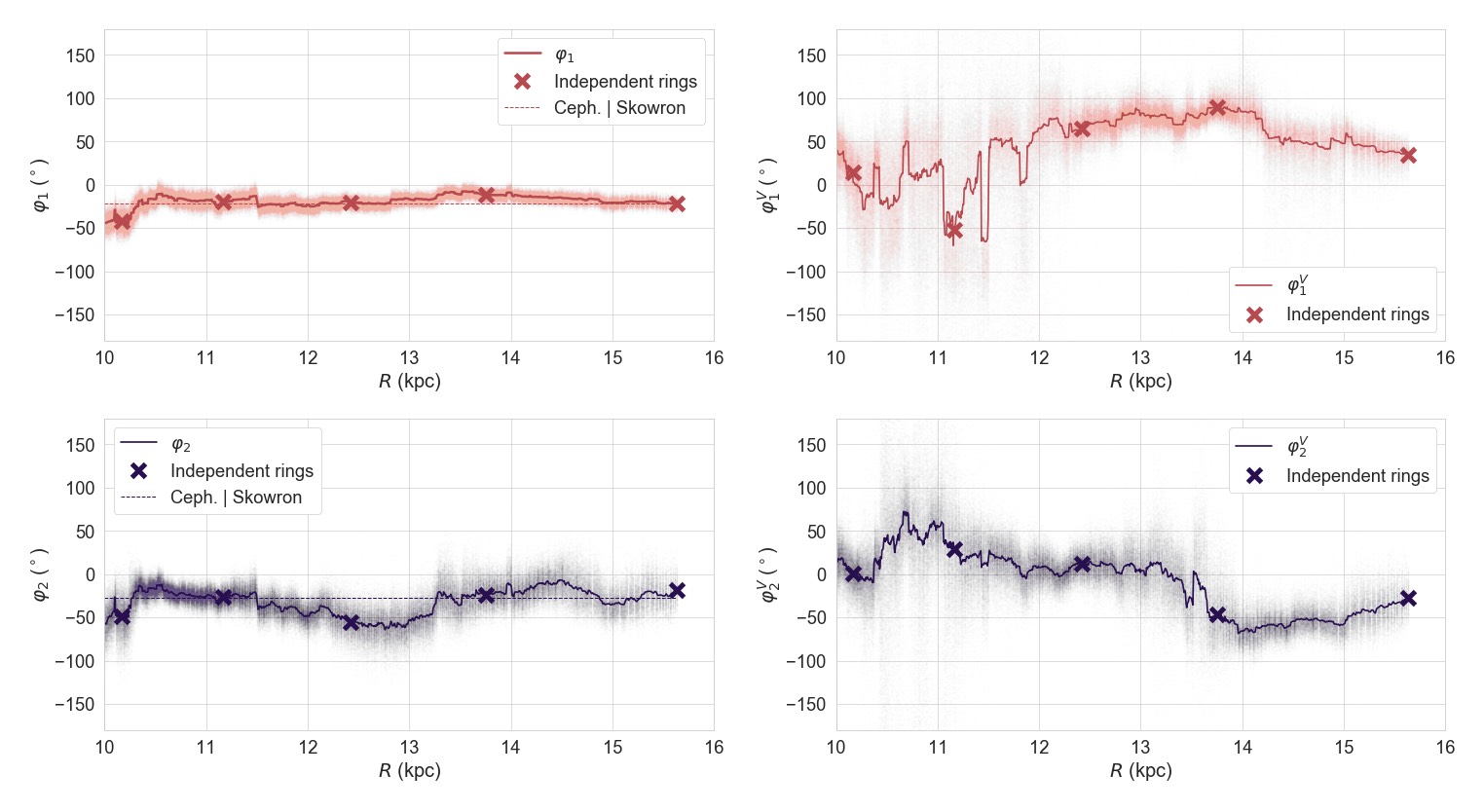}
     \caption{Each panel shows the phases of each mode as a function of galactocentric radius. The first two top panels are the results for $Z$ and the bottom two for $V_z$ (left $m=1$ and right $m=2$). The doted line for the phases in $Z$ are the constant phases obtained by \citet{Skowron2020}. The colour dots around each mode are $500$ realisation taken from the posterior at each ring.}
    \label{fig:Phi_R}
\end{figure*}

\subsubsection{Fits in $z$}

In Fig.~\ref{fig:A_R} (left panel), we present the results of the amplitudes for each mode and the intrinsic dispersion in $Z$ as a function of radius. Clearly the $m=1$ mode (red) dominates the fit (it has a maximum of $\approx 1.1$~kpc), as expected from an almost S-type like the Milky Way warp. The main mode that takes into account the asymmetries is $m=2$ (violet), its amplitude begins to grow at $R\approx 10$~kpc but never exceeds $250$~pc. For $m=0$ (yellow) we have a maximum of $\approx200$~pc. This mode can give asymmetry between both extremes of the warp, but its main purpose is to set the mean height in each ring, so it has the ability to represent radial ripples with no azimuthal dependence.
For comparison, we plot the amplitudes for each mode from \cite{Skowron2020} (dotted curves) obtained with exactly the same Cepheid sample but under the assumption of a monotonic dependency of $A_m$ with $R^2$. For the $m=1$ mode at $R>10$~kpc both amplitudes are practically the same; for the other modes the amplitudes obtained by \cite{Skowron2020} are similar to the mean behaviour of our results. 

The wavy pattern in the amplitudes for $R<10$~kpc should not be fully taken as real corrugations in the modes. In Sec.~\ref{sec:RecOfModes} we concluded that stochastic clumps in the  $\phi-z$ plane due to the SF generate correlations between the modes. This wavy pattern in $A_1$ is removed if we take $M=1$, so the wavy pattern is mainly due to correlations between $m=1$ and $m=2$. 

In Fig.~\ref{fig:Phi_R} we present the phases of the $m=1,2$ modes for $Z$ (top right for $m=1$ and top left for $m=2$) as a function of $R$. First, lets consider $\varphi_1$, our results and those of \cite{Skowron2020} (dotted line) coincide in their general trends for the external region of the disc ($>10$~kpc). For $m=1$, a twist in the direction of the galactic rotation is well defined, beginning at $R\approx13$~kpc. For the internal region both phases are difficult to determine due to the low amplitude of the warp and because the azimuthal coverage is affected by the SF. For $\varphi_2$ there is more uncertainty than for $\varphi_1$ because $m=1$ is better defined and dominates the warp. Within its uncertainty $\varphi_2$ agrees with the phase obtained by \cite{Skowron2020} (red dotted line). For $R>10$~kpc the phases, like the amplitudes, are better behaved than in the internal disc as we expected from Sec.~\ref{sec:RecOfModes}.

Finally, given that we calculate the intrinsic dispersion for $Z$ in each ring, we can see how the disc traced by Cepheids becomes thicker at larger radius, as its shown with the black curve in the left panel of Fig.~\ref{fig:A_R}. This shows how the flare in this young population starts at around $R\approx 8$~kpc with a height $\approx100$~pc to end up at a height $\approx390$~pc at $R\approx15$~kpc. Previous measurements on how thick the disc traced by Cepheid is \citep{Skowron2019,Chen3Dmap} agree with our results for the scale and trend found from $\sigma_{ID}$.

\subsubsection{Fits in $v_z$}

The right panel of Fig.~\ref{fig:A_R} presents amplitudes for the fits in vertical velocity as a function of galactocentric radius. For $V_z$ the amplitudes show a smooth oscillating pattern. The important difference between the $A_m$ and $V_m$ is that in $Z$ the $m=1$ mode dominates the warp at all radii; in $V_z$ the kinematic signal of the warp is dominated by both $m=1$ and $m=2$, a result unexpected for a tilted rings model. The $m=1$ mode in $V_z$ starts to appear at $R\approx 12$~kpc and at its maximum reaches an amplitude similar to the value of $\sigma_{ID}\approx 7.2$~\kms. 
For $m=2$ in $V_z$ there is an oscillation, as in for $m=1$ too.
The amplitude of none of the kinematic modes never exceeds the intrinsic dispersion, by contrast to the warp in $Z$, in which they do. However, the amplitude of the oscillations in $m=1,2$ is larger than the uncertainty in each mode, making the result more significant.


For the phases, $\varphi^V_1$ rises for $R>11$~kpc and $\varphi^V_2$ is nearly constant, declining for $R>14$~kpc. Since the amplitudes in $V_z$ for $m=1$ and $m=2$ are comparable, the connection between these behaviours and the twisting of the \LMV\ is not as straightforward as in $Z$ where $m=1$ clearly dominates and the LON twist is evident in the decline of $\varphi_1$ for outer radii.

Finally, the intrinsic dispersion for $V_z$ (black curve, Fig.~\ref{fig:A_R} right panel) is found to be almost constant with radius at $\sigma_{ID}\approx 7.2$~\kms. 

\bsp	
\label{lastpage}
\end{document}